\documentclass[format=acmsmall, review=False, screen=true]{acmart}

\definecolor{highlight}{RGB}{0, 0, 200}

\acmJournal{TOIS}
\acmVolume{9}
\acmNumber{4}
\acmArticle{29}
\acmYear{2021}
\acmMonth{6}
\copyrightyear{2021}

\setcopyright{acmlicensed}

\acmDOI{0000001.0000001}

\usepackage{float}
\usepackage{multirow}

\begin{document}
\setcopyright{acmcopyright}
\acmJournal{TOIS}
\acmYear{2021} \acmVolume{1} \acmNumber{1} \acmArticle{1} \acmMonth{1} \acmPrice{15.00}\acmDOI{10.1145/3507357}

\title{Simulating and Modeling the Risk of Conversational Search} 


\author{Zhenduo Wang}
\email{zhenduow@cs.utah.edu}
\orcid{1234-5678-9012}
\affiliation{%
  \institution{University of Utah}
  \city{Salt Lake City}
  \state{Utah}
  \country{United States}
  \postcode{84112}
}

\author{Qingyao Ai}
\email{aiqy@cs.utah.edu}
\affiliation{%
  \institution{University of Utah}
  \city{Salt Lake City}
  \state{Utah}
  \country{United States}
  }
  
\renewcommand{\shortauthors}{Wang and Ai, et al.}

\begin{abstract}
    In conversational search, agents can interact with users by asking clarifying questions to increase their chance to find better results. Many recent works and shared tasks in both NLP and IR communities have focused on identifying the need of asking clarifying questions and methodologies of generating them. These works assume asking a clarifying question is a safe alternative to retrieving results. As existing conversational search models are far from perfect, it’s possible and common that they could retrieve/generate bad clarifying questions. Asking too many clarifying questions can also drain user’s patience when the user prefers searching efficiency over correctness. Hence, these models can get backfired and harm users' search experience because of these risks by asking clarifying questions. 

    In this work, we propose a simulation framework to simulate the risk of asking questions in conversational search and further revise a risk-aware conversational search model to control the risk. We show the model’s robustness and effectiveness through extensive experiments on three conversational datasets, including MSDialog, Ubuntu Dialog Corpus, and Opendialkg in which we compare it with multiple baselines. We show that the risk-control module can work with two different re-ranker models and outperform all the baselines in most of our experiments.
\end{abstract}




\maketitle

\section{Introduction}

Users often cannot express what they need effectively to modern IR systems, and it has become an important reason for the retrieval of unreliable results in practice. To address this problem, a diversity of techniques and UI design has been proposed to help users form high-quality information requests, such as query auto-complete \cite{autocomplete}, query suggestion \cite{querysuggestion1, querysuggestion2, querysuggestion3, querysuggestion4}, etc. In particular, conversational retrieval that actively asks clarifying questions to improve information request quality has been widely believed to be the key technique for future IR \cite{framework2017}. There are many recent studies and shared tasks \cite{whatdoyoumean, cqidentify, raodaume2019, xuasking2019, askingcq,zamania} \cite{trechard,aliannejadi2020convai3} on identifying clarifying question needs and retrieving or generating good clarifying questions. The motivation is that the additional information in user responses to the clarifying questions can refine the initial information requests and guide conversational retrieval.

An often ignored possibility of existing studies on conversational retrieval is that conversational search paradigms themselves could contain false assumptions and bring risk to users and modern IR systems. Previous work on this topic assumes that users are dedicated to search sessions after they submit an information request and would provide responses to any questions asked by the system \cite{aorb, systemaskuserrespond, conversationalrecommend, negativefeedback}. Such over-optimistic assumption neglects the risk of asking an inappropriate question to users. For example, asking a question with racial innuendo [33] could offend a user; asking too many over-specific questions could also could overwhelm a user, etc. Apart from the negative impressions brought to users by bad clarifying questions, conversational retrieval systems can also get noisy responses from users by asking questions that have unclear intents, and can eventually provide low-quality retrieval results. All the above cases will harm users' experience and potentially drive users away from using the IR system again. Hence, asking clarifying questions is a risky decision in conversational search. 

In this paper, we propose a novel risk evaluation process based on a user simulation experiment. In the user simulation experiment, the tested conversational search system needs to interact with the user based on query and conversation context in multiple rounds and decide between returning search results and asking clarifying questions. The simulation experiment aims to test whether the conversational search system can finally retrieve relevant results for the user’s initial information need through multi-turn conversation instead of the accuracy of response selection/re-ranking in a single turn. The user simulator can respond to the system’s clarifying questions by using answers from the dataset or choosing to leave the conversation because of bad clarifying questions. We include multiple evaluation metrics to evaluate the system’s performance from different aspects comprehensively, including result reliability and system decision accuracy. 

\begin{figure}[h]
  \centering
  \includegraphics[width=0.7\linewidth]{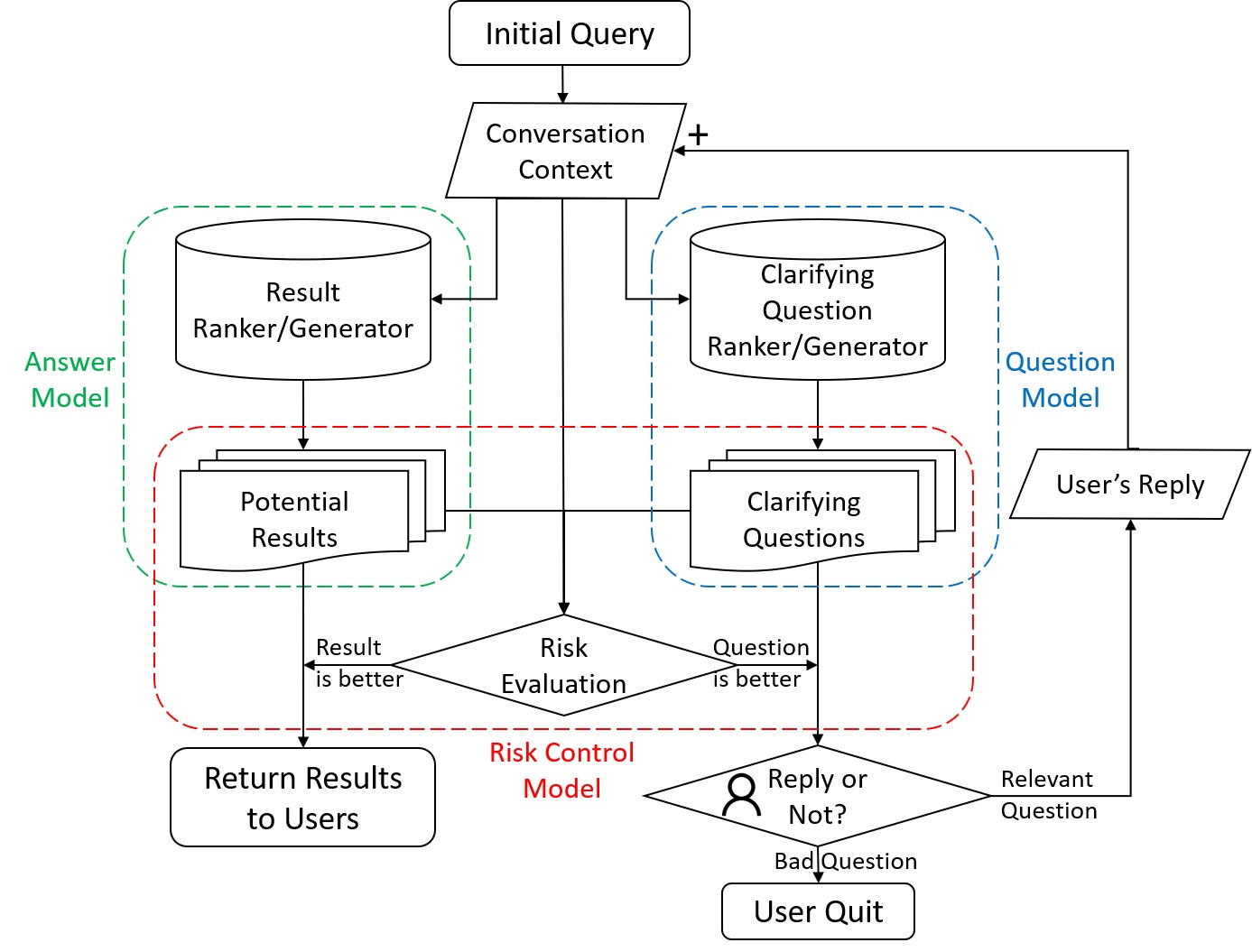}
  \caption{Risk-aware Conversational Search Framework}
  \Description{Risk-aware Conversational Search Framework}
  \label{riskawareframework}
\end{figure}

We also revise our previous risk-aware conversational search framework \cite{wang2021controlling} which aims to control and balance the risk and cost of result reliability and user engagements in modern IR systems. The risk-aware system comprises three components as shown in Figure~\ref{riskawareframework}: (1) A result retrieval model that retrieves and ranks candidate results according to the current information need and corresponding conversation context; (2) A question retrieval model, similar to those in previous studies \cite{msdialogrank, askingcq}, that retrieves or generates clarifying questions for users, hoping that user’s responses could enrich the context and further improve the quality and credibility of results provided by the result retrieval model; And (3) a risk control model which decides whether the system should ask the clarifying question provided by the question retrieval model or directly show the results provided by retrieval model. Counterparts of such a decision-making model have been seen in previous works regarding clarifying questions like \cite{xuasking2019, aliannejadi2020convai3}. However, their works try to predict whether asking clarifying questions is needed in a specific context. In contrast, our risk control model decides whether to ask a question according to both the context and the possible candidate questions retrieved by the retrieval models. The difference is that the revised decision-making model considers both the clarifying question need and the retrieved questions qualities. The risk control model is the key for this framework to control and balance the risk of asking bad clarifying questions or providing immature results to users. 

In our previous work, we only test the framework on one dataset and we assume the only difference in users is the tolerance for bad clarifying questions. In this paper, we test this framework on more diversified datasets and user types and show that our framework can generalize to them. To be more specific, we build two different risk-aware conversational search agents by combining two different state-of-the-art result and question retrieval models with our risk control model that is trained using reinforcement learning. We conduct various simulation experiments on three different conversational datasets and with three different user simulator models. For comparison, we also implement several baseline conversational search strategies, including (1) always directly answer the query, (2) always ask clarifying questions before answering the query, and (3) predict whether to ask clarifying questions based on conversation history. Through simulation experiments, we show that the risk-aware conversational search agent outperforms several baseline conversational search strategies in most of our experiments. We also provide new insights and explanations of how our framework improves the search result quality and user experience through result analyses and case studies.

We consider our contributions in this paper as follows: 
\begin{enumerate}
    \item We propose a conversational search risk simulation experiment to evaluate conversational search systems. The simulation experiment can better evaluate whether conversational search agents can improve answer quality and user experience at the whole-conversation level. We also propose to use three different evaluation metrics to evaluate the simulation experiment results.
    
    \item We revise our previous risk-aware conversational search framework. Through extensive simulation experiments, we test the framework on different datasets and let it interact with different user-type simulators. We show that our previous conversational search framework can generalize well on multiple datasets and user types. We also provide insights into these improvements through analyses and case studies.
    
We have published our code and datasets on GitHub.\footnote{https://github.com/zhenduow/conversationalQA} 
\end{enumerate}

\section{Related Works}
\subsection{\textbf{Conversational search}}
Many previous works have studied the problem of the vagueness of user queries in conversational search. One approach is to analyze the links among conversation turns. Kaiser et al. \cite{wordproximitynetwork} build a word proximity network to compute word coherence besides query-response relevance. Aliannejadi et al. in \cite{relevantturn} estimate relevance between current and previous turns and incorporate it in context representations. Another approach is to alter or enhance the original query by query suggestion or query auto-completion. The former provides the user with several possible clearer queries that are similar to the user's query. Boldi \cite{boldi2008, boldi2009} uses query flow graph for query suggestion. Rosset et al. \cite{rosset2020leading} propose to rank query suggestions by measuring their relevance and usefulness. Compared to query suggestion methods, query auto-completion emphasizes more on additive changes. Voskarides et al. in \cite{queryresolution} resolve anaphoras in later conversation turns by traversing through all previous turn terms and determining whether to add each of them to the current query. Bar et al. propose to improve query auto-completion results by incorporating contexts \cite{querycompletionwithcontext}.

Conversational search is usually considered different due to the complexity of conversation structures. Work such as \cite{yangnextquestion, topicpropagation, qu2020open} studies conversations in a session where users can ask multiple questions potentially with multiple topics. We consider this as fundamentally different from the conversations we are interested in. The conversational search problems we are working on are entire conversation is centered on the user's initial information need.

\subsection{\textbf{Asking clarification questions}} Another widely studied direction of resolving query vagueness is to ask clarifying questions. A considerable amount of attention has been put on asking clarification questions in conversational search by natural language processing and information retrieval community. Earlier in the TREC 2004 HARD track\cite{trechard}, asking clarifications was permitted for participants to get additional information. Rao and Daumé III in \cite{raodaume2018} propose a clarification question selection model which maximizes the expected value of perfect information after getting the answer to the clarification question. Later in \cite{raodaume2019}, the model is extended to a reinforcement learning model which better suits multi-turn conversation scenarios. Recent works approach asking clarification questions in various ways. Aliannejadi et al. \cite{askingcq} select clarifying questions from a collection of manually generated questions. Zamani et al. \cite{zamania} create a clarifying question taxonomy based on user studies and automatically generate clarification questions via multiple approaches and \cite{zamani2020engagement} analyze the user engagements received from clarifying questions. Later, Zamani et al. \cite{zamani2020mimics} publish an annotated dataset comprising real-world user interactions. Cho et al. \cite{askingcqnlgmulti} generate one question that can capture common concepts in multiple documents. 

While methodologies of retrieving or generating clarification questions have been studied by these previous works, some other recent works also study when to ask clarification questions, i.e., identifying the need of asking clarifying questions. Xu et al. \cite{xuasking2019} create a clarification identification and generation dataset where clarification identification is cast as a binary classification problem. Most recently, Aliannejadi et al. \cite{aliannejadi2020convai3} organize a shared task that poses questions on when to ask clarifying questions and how to generate them, concluding all the aforementioned works. 

\subsection{\textbf{Risk control}}
Radlinski and Craswell in \cite{framework2017} propose a theoretical framework for conversational search, where they define action spaces for both the user and the agent and emphasize the necessity of a model which can decide among the actions. Few existing works have addressed the conversational search risks in making these decisions, which is a key factor in real-world IR applications. Su et al. \cite{surisk} propose a risk control framework for web question answering. They use an uncertainty model, which estimates the predictive uncertainty, and a decision model, which decides whether to answer the question, to control the risk of their reading comprehension system. Our system extends their decision model by allowing the agent to ask clarification questions instead of doing nothing when the query is not answerable. We simplify the action space of the agent into either answering the query or asking a clarification question inspired by the conversational rules described in \cite{emanuel1968sequencing} that user's questions are followed by either answer or clarifying questions in-between. Wang et al. \cite{wang2021controlling} propose a risk controlling framework that can decide between returning the results and asking clarifying questions in conversational search and show it can improve multiple baseline models when interacting with different user types with different tolerances of irrelevant clarifying questions. Our work generates their experiments on more datasets and more user settings with different patience for clarifying questions. In addition, we provide insights into their framework through analytical studies and case studies.

\subsection{\textbf{Datasets}}
We use three datasets in this work, MSDialog, Ubuntu Dialog Corpus \cite{udcdataset}, and Opendialkg \cite{opendialkgdataset}. In \cite{msdialogintent, msdialogrank, msdialogintent2}, Qu and Liu introduce MSDialog dataset that comprises conversational question answering thread collected from Microsoft forum. A large number of questions fall into one of the major Microsoft products, such as Windows, Office, and Bing. In their works, they perform two tasks on the dataset. Qu \cite{msdialogintent, msdialogintent2} analyzes and characterizes users' intent of each utterance in the conversation, in which they include clarification questions as one type of intent. Liu \cite{msdialogrank} uses the conversation utterance pairs to train a response re-ranker, which can rank response candidates given the current contexts. Ubuntu Dialog Corpus dataset \cite{udcdataset} contains logs from Ubuntu-related chat rooms and is also usually used as a response ranking dataset. The Opendialkg dataset \cite{opendialkgdataset} contains conversations where a user asks an agent for a recommendation or opinion about a certain movie, music, books, etc. In \cite{opendialkgdataset}, Moon et al. propose an entity prediction task where the model retrieves a set of relevant entities given multiple modalities of conversation contexts, including conversation history texts and knowledge graph entities. Our task differs from all of them in that our task simulates the whole conversation process iteratively via the interactions between user and agent. 

\section{Risk Modeling and Simulation in Conversational Search}

In this section, we propose a novel risk evaluation process based on user simulation. We include multiple evaluation metrics which cover different aspects of the experiment from both the result and system perspective.

\subsection{Problem Formulation}

Our work proposes a simulation framework for the conversational search problem. Before talking about our solution, we formally describe our formulation of the conversational search problem. 

In conversational search, users usually start the conversation with an initial information request query $q$ with or without further description as context $h$. A conversational search system aims to provide a final answer $a$ to the query through retrieval or answer generation. Before providing the final answer, the system can ask a certain number of clarifying questions $cq$ in order to seek more information from the user. However, asking clarifying questions can lead to unpredictable outcomes depending on the user. When the user considers the clarifying question relevant to the query and the context, the user is usually willing to answer the question and provide more information. When the user considers the clarifying question as bad and not worthy to answer, the user will usually leave the conversation. Hence, the number of clarifying questions that the system will ask depends on not only its recognition of the information completeness, but also the user's patience to answer these questions.(which is unknown to the system). The goal of a conversational search system is to maximize the quality of the final answer while still keeping the user engaged in the conversation.

More formally, we model users' willingness to interact with the conversational search system with two parameters. The first parameter is the user's tolerance $\tau$ to bad clarifying questions. As described earlier, the user will usually lose their faith in the conversation system when seeing bad clarifying questions but not necessarily \cite{zamania}. Some users are more tolerant and they will stay for just a few more turns for better responses from the system. Here, we set $\tau$ as the number of bad clarifying questions that the user can tolerate before leaving the conversation. The second parameter is the user's patience $\rho$ for questions overall. We believe that users have different expectations for conversational search efficiency, i.e., the number of questions to be asked before finally being returned the answer. For example, a user that asks for a recommendation for a movie may not want to answer a long list of clarifying questions about preferences. Here, we set $\rho$ as the number of questions that the user will answer regardless of the question quality.

With the above definition, the risk-aware conversational search problem can be formalized to maximize the answer quality by asking less than $\rho$ clarifying questions, and at most $\tau$ bad clarifying questions. We explain how to evaluate the answer quality later.

\subsection{Evaluation based on User Simulation}
In this section, we introduce our user simulation framework for conversational search and evaluation. 

Assume the dataset $D=\{D_1,...,D_M\}$ contains multi-turn conversations which can be represented as $D_{i}=\{u_1,a_1, ...,u_n,a_n\}$, where $u_1,...,u_n$ are user utterances and $u_1$ is the initial query, and $a_1,...,a_n$ are conversational search agent utterances. $a_1,...,a_{n-1}$ are clarifying questions, and $a_n$ is the final answer to the query. We assume that $u_2$ is the user's answer to agent's clarifying question $a_1$, $u_3$ is the answer to $a_2$, and so on. For a conversation context $\{a_1, u_1, ..., a_i\}$, the positive clarifying questions are all the clarifying questions in the original conversation after $a_i$, i.e., $\{a_{i+1},...,a_{n-1}\}$, the only positive answer to the query is $a_n$. We will introduce the datasets we use in details and provide examples in Section 5.1.

To evaluate a conversational search system's performance at the entire conversation level, we design a user-system simulation experiment to mimic the user-system interactions in multiple turns instead of evaluating the next response prediction in a single turn. With the dataset definition above, starting from the user's initial query $D_i=\{u_1\}$, the system is assumed to returning an answer or asking a clarifying question in each round. The user simulator will respond to the agent response in the following steps: 
\begin{enumerate}

    

    \item Starting from the first round and in each round, the conversational search system decides whether to answer the query or ask a clarifying question. Then it responds with the answer/question response.
    
        \item If the system returns an answer in (1), the simulation for this conversation will end. The user simulator will evaluate the answer. 
        
        \item If the system asks a question in (1), then the user simulator compares the total number of asked questions with the user's patience for questions $\rho$. If more than $\rho$, the simulation will end. In this case, the user simulator will record a failure on this conversation. If less than $\rho$ and the question is relevant, then the user simulator will reply with feedback. In a conversation $\{u_1,a_1,u_2,a_2, ...,u_n,a_n\}$, if the clarifying question is $a_j,(1\leq j\leq n-1)$, then user will reply with $u_{j+1}$. Both the clarifying question $a_j$ and the user's feedback $u_{j+1}$ will be appended to conversation context ($D_i=D_i + \{a_j,u_{j+1}\}$). Then repeat step (1). When the system goes back to (1), it may not repeat the same clarifying questions in the future. We also add two rules for determining relevant/irrelevant clarifying questions: (1) A clarifying question that originally appears in later turns of the conversation is still a relevant clarifying question, i.e., in $\{u_1, a_3\}$, $a_3$ is a valid clarifying question to $u_1$; (2) Also, firstly asking a clarifying question that originally appears later in the conversation, and then asking a clarifying question that originally appears early is fine, i.e., $\{u_1, a_3, u_3, a_1,...\}$ is a valid conversation sequence. These are from an empirical assumption that the conversations are order-independent on the datasets we use. 
        
        \item If the question is irrelevant, the simulation of this conversation will not necessarily end. As explained earlier, users are not necessarily depressed by irrelevant clarifying questions. If the user has already seen $\tau$ bad clarifying questions, the simulation will end. In this case, we will record a failure in this conversation. If the user is not out of tolerance $\tau$, the agent will redo steps (1) and know not to ask the irrelevant questions again. 
\end{enumerate}

By definition, the simulation experiment is guaranteed to end with either returning an answer to the user or causing the user to leave in no more than $n$ turns, where $n$ is the total length of the original conversation. After the simulation experiment ends, we will evaluate the system's output based on how the simulation ends with the following evaluation metrics.

\subsection{Evaluation Metrics}

To evaluate our simulation experiment results, we can either just evaluate the final answer quality or count in the systems' decision-making process. In our evaluation, we include three metrics: Recall@1, Mean Reciprocal Rank, and decision error rate. Each of them measures the answer quality and user experience or model performance from a different aspect. Hence, we include all of them to better evaluate them and understand their strengths and weaknesses.

\subsubsection{\textbf{Recall@1}}
The Recall@1 metric is defined as the frequency of conversations where users finally receive a relevant answer, regardless of how many relevant or even irrelevant clarifying questions have been asked during the process. As discussed previously, the simulation can end with two general cases, the user receives an answer or the user leaves the conversation. When the user receives an answer, Recall@1 is 1 if the answer is the relevant answer, otherwise 0. When the user leaves the conversation, Recall@1 is 0. We consider recall@1 more of an evaluation metric for answer quality and user experience rather than models' decision-making progress.

\subsubsection{\textbf{Mean Reciprocal Rank}}
Similar to the Recall@1 metric, Mean Reciprocal Rank (MRR) is also a metric defined on the final answer. While the recall@1 metric only cares about the answer returned to the user, MRR cares about the entire re-ranked answer list. In recall@1, there is only 0 or 1 for a conversation, but in MRR, a conversation can get partial credit as long as the correct answer ranked among the top10, i.e., if a relevant answer is ranked second by the answer re-ranker, then that conversation will receive a 0.5 as reciprocal rank. For conversations where the relevant answer is ranked outside top10 or users leave before receiving the answer, RR is still 0. 

Compared to recall@1, MRR gives answers partial scores, hence models which frequently ask irrelevant clarifying questions will be punished more in this metric. This is because they often can get non-zero scores if they choose not to ask those irrelevant questions. In this sense, the MRR metric is not a metric purely for evaluating the answer quality. It is a metric that combines answer quality and models' decision-making progress.

This metric can only be used when the answer/question retrieval models are ranking models.

\subsubsection{\textbf{Decision error rate}}

The decision error rate metric is defined as is the number of worse decisions divided by the number of total decisions. Worse decisions are those decisions that lead or can lead to lower MRR of the conversation, e.g., asking an irrelevant clarifying question which results in 0 MRR or answering with suboptimal answers which miss the chance of improving MRR by asking clarifying questions. We define answers with reciprocal rank lower than $1/\tau$ (i.e., rank $\tau$ or lower) as suboptimal answers, using the same $\tau$ as user's tolerance for irrelevant clarifying questions. The intuition behind this definition is that users need to spend the same amount of tolerance to check the top $\tau$ answers as checking $\tau$ irrelevant clarifying questions. However, the latter is more likely to improve the conversation MRR in future turns.

Compared to recall@1 and MRR, this metric mainly measures the decision-making progress, which is an underlying factor that is hidden from the final output during the simulation experiment. This is purely a system-side metric. It may not reflect the answer quality and user happiness the best, but we consider it very helpful in understanding the decision-making progress and how much our risk control model improves over the baselines. 

There is also one notable difference between decision error rate and both recall@1 and MRR. Recall@1 and MRR are computed per conversation, while decision error rate is computed per decision during the simulation experiment. Agents can make several decisions in one conversation, and the number of decisions in a conversation is varying. This implies that each conversation contributes unevenly to the decision error rate, which also makes it less suitable for evaluating answer quality and user experience. 

This metric only measures the risk control decision-making part, thus it is extendable if we want to combine our risk control model with other answer/question generation models.

\subsection{Caveats and limitations}

We make several assumptions and approximations in our simulation experiment and evaluation. They are made based on our observations and understandings of the datasets. They simplify our experiment but can also set limitations to our experiment and conclusions.

The first approximation is that we sample negative clarifying questions and search answer candidates from other conversation threads. This is simple but can be less effective. One reason is that the sampled questions and answers can be relevant to the search context. The other reason is that the sampled negative candidates only have features within the dataset, like MSDialog and Ubuntu Dialog Corpus, which are limited in topics. In a real-world search scenario, the negative candidates may come from all types of corpora, and the re-rankers need to rank all of these candidates.

We also assume that all clarifying questions within the same conversation thread are valid questions and clarifying questions are order-independent. This assumption is reasonable in MSDialog and Ubuntu Dialog Corpus because user queries in these datasets are usually talking about issues in machines and clarifying questions are mostly asking about system configurations in the user's machines. The clarifying questions are mostly agents asking about system configurations. Hence, the order rarely matters in these conversations. However, these are still rather strong assumptions and set limitations to our experiments and conclusions.

Another limitation of our experiments is that we only assume one user type in each experiment, which is unrealistic in real-world search scenarios where a search system must face different user types at the same time. This is because optimizing the performance for one user type at a time differs from optimizing the performance for a mix of user types. It can be a promising future direction for developing new evaluation metrics that are compatible with different user types or based on more flexible user settings.

\section{Risk-aware Conversational Search Model}
Based on our problem formulation, we now introduce our previous risk-aware conversational search model, which can be divided into 3 parts, including the two retrieval models and a risk control model. The answer retrieval model reads the conversation contexts and chooses the best answer to the query. The question retrieval model does similar for clarifying questions. Then the risk control model reads the conversation contexts together with the retrieval models' outputs and evaluates the two action choices. If the decision is to answer the query, then the conversational search will end with the agent providing the top retrieved answer. If the decision is to ask a clarifying question, then the agent will ask the top retrieved question to the user and collect feedback from the user, which may give the agent more information to retrieve better answers in future turns.

\subsection{Answer and question retrieval model}
The goal of answer and question retrieval models is to retrieve the best answer or clarifying question given the query and current conversation contexts. In general, the retrieval task is in the form of $R(q) = p$, where $q$ is the conversational search context and $p$ is the retrieved response. It can be further defined as different tasks, including response candidates re-ranking, response generation, or other tasks, as long as the task output is a text response to the conversational search state. In this work, we use the response re-ranking tasks and models as an example. Our answer re-ranking model aims to re-rank all the answer candidates given the current conversation context, and our question re-ranking model aims to re-rank the clarifying questions.

To demonstrate that our risk control model can cooperate with any answer/question re-ranker model, we test it with different re-ranker models. The re-ranker models we test are the ParlAI bi-encoder and poly-encoder re-rankers \cite{polyencoder}. In each experiment, we use the same re-ranker structure for the answer and question re-ranker models, but with independently trained model parameters.

\subsubsection{\textbf{Bi-encoder}}

The bi-encoder re-ranker uses two pre-trained transformer \cite{bert} encoders $E_q$ and $E_p$ to encode the conversation context $q$ and response candidates $P=\{p_1,...,p_n\}$ separately, where the response candidates are either answer candidates or question candidates depending on the ranking task. The transformer encoder takes a sequence of natural language text as input and outputs a sequence of encoding vectors of the same length as the input sequence. Each vector represents an input word correspondingly. Before encoding, a special word token `[S]' is appended to the beginning of both conversation context and each candidate. After encoding, the representation of `[S]' is used as the representation of the entire input sequence. The encoded context representation $E_q(q)$ and candidate representations $\{E_p(p_1), ..., E_p(p_n)\}$ are used for scoring. Each candidate score is computed as the dot product between the context vector and candidate vector. Finally, all the candidates are ranked by their candidate scores. The entire bi-encoder model can be represented as:

\begin{equation}
s_{i} = E_p(p_i)\cdot E_q(q)
\end{equation}

\subsubsection{\textbf{Poly-encoder}}

An issue in the bi-encoder is that the conversation context and candidates cannot see each other during encoding. There is only self-attention inside conversation context and candidates. In poly-encoder, this issue gets addressed. A structurally simple way to get access to context-candidate is to concatenate them and pass them into a single encoder. This structure computes the full co-attention between context and candidates and is referred to as cross-encoder \cite{polyencoder}. However, this structure has significantly higher computational complexity. Poly-encoder's structure almost approximates full co-attention without loss of efficiency. Similar to the bi-encoder, the poly-encoder also uses two pre-trained transformer encoders to encode the conversation context and candidates separately. When computing context-candidate attentions, poly-encoder computes $m$ code-attended representations for the context instead of the original sequence of $N$ vector representations, where $m$ is a hyper-parameter that affects encoding speed and $N$ is the context length (typically $m<N$). Usually, conversation context is much longer than candidates, thus it is the primary cause of slowing down the attention computations. Reducing the $N$ vectors sequence to $m$ vectors can significantly reduce the attention complexity. The $m$ representations are computed using the outputs from the last layer of the context encoder:

\begin{equation}
q_{\text{attn}}^{C_i} = \sum_j w_j^{C_i}h_j 
\end{equation}
where $(w_1^{C_i},..., w_N^{C_i}) = \text{softmax}(C_i\cdot h_1, ..., C_i\cdot h_N)$ and $(h_1, ..., h_N)$ are last layer context encoder outputs. The codes $(C_1, ..., C_m)$ are randomly initialized and optimized during training. The $m$ code-attended context representations are then used to compute the final candidate-attended context representation:

\begin{equation}
q_{\text{attn}} = \sum_i w_iq_{\text{attn}}^{C_i}
\end{equation}
where $(w_1,..., w_m) = \text{softmax}(E_p(p)\cdot q_{\text{attn}}^{C_1}, ..., E_p(p)\cdot q_{\text{attn}}^{C_m})$.

The candidate representations are still computed by the candidate encoder alone, just like in bi-encoder. With the attended context representation and the candidate representation, candidate scores are computed similarly as in bi-encoder:

\begin{equation}
s_{i} = E_p(p_i)\cdot q_{\text{attn}}
\end{equation}

Poly-encoder structure simplifies the necessary but expensive co-attention computations and combines the best of both bi-encoder and cross-encoder. It can achieve almost the same performance as cross-encoder in experiments \cite{polyencoder}.

\subsection{Risk Control Model}

Our risk control model reads the current conversation context, and the outputs from the answer and question retrieval models to decide which is the better choice considering both the conversation context and retrieval models' capability. The risk control model is essentially a deep Q network (DQN). Its input is a list of text and numerical features, and its output is a $2x1$ vector $y_{\text{pred}} = (r_{ans}, r_{cq})$ denoting the expected rewards from taking the two actions, namely answering the query with the top retrieved answer and asking the top retrieved clarifying question.

\begin{figure}[h]
  \centering
  \includegraphics[width=0.7\linewidth]{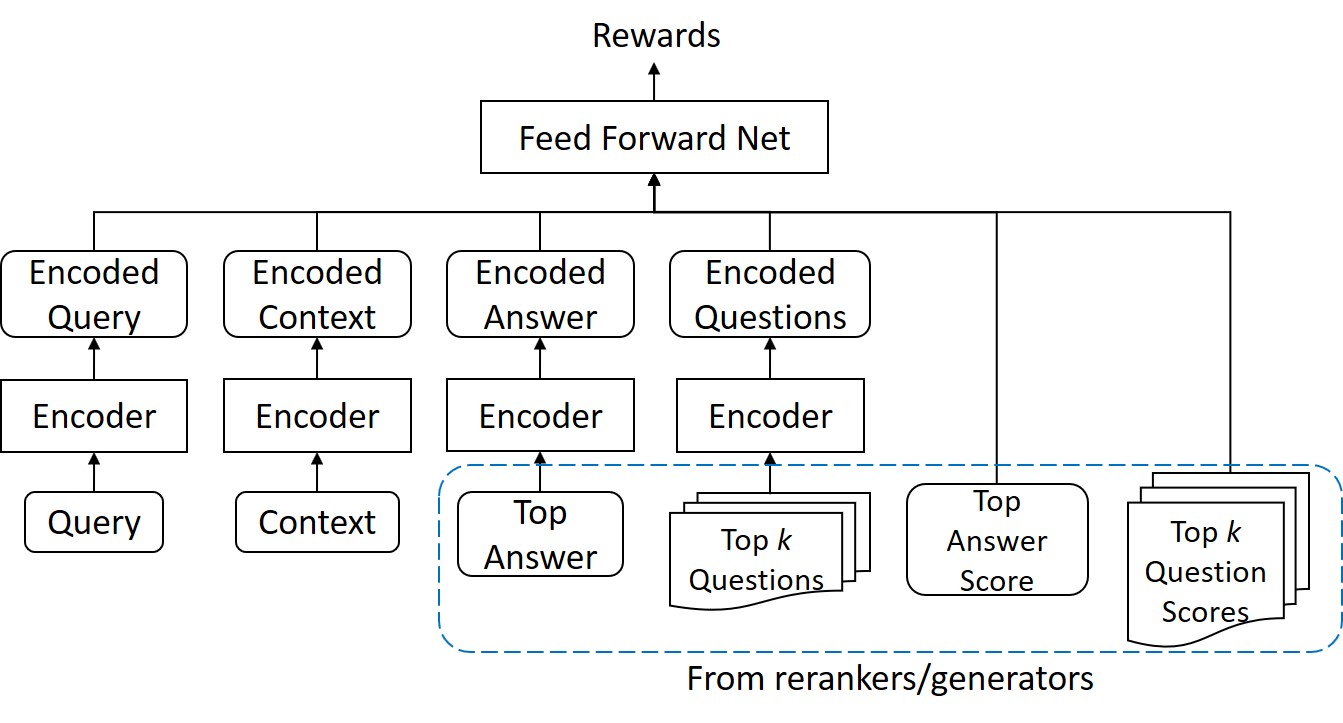}
  \caption{Risk Control Deep Q Network}
  \Description{Risk Control Deep Q Network}
  \label{DQN}
\end{figure}

The input features of the risk control DQN are the initial query $Qu$, conversational history $H$, the top retrieved answer $a_1$ and top $k$ clarifying questions $\{cq_1,...,cq_k\}$, the retrieved answer ranking score $s_{ans}^{1}$, and the retrieved question ranking scores and  $s_{cq}^{1:k}$. The text features are separately encoded using transformer encoders and the vectors for [CLS] token are used as representations for each text feature. Then the DQN concatenates all the features like a state vector, which can be represented as $S = (q, h, a^1, cq^1,..., cq^k, s_{ans}^{1}, s_{cq}^{1:k})$. The DQN is essentially a 2-layer feed-forward neural network:

\begin{equation}
D(S) = W_2\cdot \phi(W_1\cdot S + b_1) + b_2
\end{equation}
where $W_1, b_1, W_2, b_2$ are learnable parameters and $\phi$ is ReLU activation function. Note that the DQN does not have an activation function in the second layer since it aims to predict reward values instead of classification probabilities. The structure of the DQN is shown in Fig.\ref{DQN}. We also do ablation studies on these features, which involve two variations of our model. One uses only the textual features $(q, h, a^1, cq^1,..., cq^k)$ and the other only uses the score features $(s_{ans}^{1}, s_{cq}^{1:k})$. We will show comparison results of these model variations in the result section.

Although the structure of DQN is very simple, we do not have annotated data to supervise the training of the DQN. To overcome this challenge, we use reinforcement learning, which trains the network by simulating the interaction between agent and user. We define reward and penalty for each interaction outcome using simple settings. The DQN is trained by learning from these rewards or penalties. We will describe this in the next section.

\subsection{Training}
\subsubsection{\textbf{re-rankers}}
Before running the main simulation experiment, we first pre-train the answer and question re-rankers. Before the pre-training, we initialize the re-ranker parameters with the checkpoints provided by ParlAI which was previously trained on Wikipedia and Toronto Books and then tuned on huge Reddit dataset \cite{redditdataset}. We then fine-tune the answer re-ranker and question re-ranker respectively on answer re-ranking and question re-ranking pairs on our datasets. Assume $B$ is a batch of size $K$ in dataset $D$, $B=\{(q_1,p_1),...,(q_K,p_K)\}$ where $q$ is conversation context and $p$ is the true response. For conversation context $q_j$, the true relevance scores vector is $T_j = (t_1 = 0,...,t_j=1,...,t_K = 0)$. And the computed ranking scores vector is $S_j = (s_{j1},...,s_{jK})$ where $s_{ji} = E_p(p_i)\cdot E_q(q_j)$ for bi-encoder and $s_{ji} = E_p(p_i)\cdot q_{\text{attn}_j}$ for poly-encoder. The fine-tuning goal is to minimize the average cross-entropy loss of the computed candidate ranking scores and the true binary relevance scores in batches:

\begin{equation}
\min \sum_{B\in D} \sum_{j=1}^K H(S_j, T_j)
\end{equation}
where $H(S,T)$ is the cross-entropy loss. The dataset is randomly split into batches, thus we assume that only the true response $p_j$ is relevant to a context $q_j$ in a batch. All the other responses in the batch are used as negative samples.

\subsubsection{\textbf{Risk Control Model}}
After the re-rankers are trained, they are fixed during the risk control model reinforcement learning. We use reinforcement learning to train our DQN without supervision. At the start of training, the risk control model will randomly choose to answer the query or ask a clarifying question with the top retrieved answer and question. Any decision it makes will cause a reward or penalty, and the decisions will also affect whether the simulation of conversation continues or ends. Then it learns from the reward or penalty from these actions and gradually reduces the chance of taking random action and decides from what it has already learned. 

The rewards and penalty of each action are defined as follows. If the action is answering the query (ANS), then DQN will receive a reward $r_{ans}$ equal to the retrieved answer reciprocal rank. Recall that the input to risk control model is $S = (q, h, cq^1,..., cq^k, a^1,...,a^k, s_{cq}^{1:k}, s_{ans}^{1:k})$, and the retrieved answer part is a ranked list $a^1,...,a^k$. Thus the reward $r_{ans}\in [0,1]$, i.e., $r_{ans} = 1$ if $a_1$ is relevant, $r_{ans}=0$ if $a^1,...,a^k$ are all irrelevant, and $0<r_{ans}<1$ if a non-top retrieved answer $a^j (j>1)$ is relevant. Answering the query will always end the conversation, as it is an attempt to draw a conclusion for the user's information need. 

If the action is asking a clarifying question (CQ), then DQN will receive either a reward or penalty depending on whether the top and only the top retrieved clarifying question $cq_1$ is relevant. If $cq_1$ is relevant, then DQN will receive an immediate reward $r_{cq}$ for asking a relevant question. Otherwise, it will receive a negative penalty $p_{cq}$, because this is the risk scenario we want to avoid in conversational search. Asking a relevant clarifying question to the user will receive the user's feedback on the asked question. Then the clarifying question and the user's feedback will be appended to the conversation context and the simulation experiment will continue to a new round with updated input. Asking an irrelevant clarifying question will consume the user's patience. The agent can then remove the irrelevant question from question candidates and try again. But once the user's patience is gone, the conversation will end and the user will leave with a negative impression of the agent. 

\begin{table}[htbb]
\caption{Risk Control DQN Policy Table}
\begin{center}
\begin{tabular}{|c|c|c|}
\hline
& Relevant & Irrelevant \\\hline
Answer & \multicolumn{2}{c|}{Answer Reciprocal Rank} \\\hline
Ask & $r_{cq}$ & $p_{cq}$ \\\hline
\end{tabular}
\label{DQN Policy table}
\end{center}
\end{table}

In Table \ref{DQN Policy table}, we summarize the reward and penalty for the two actions. Because asking a relevant clarifying question will continue the simulation experiment, the immediate reward $r_{cq}$ is designed to be positive but small. Hence that DQN will not learn to accumulate $r_{cq}$ and never really answer the query. The logic behind the simulation experiment and this reward table is that we want the DQN to learn that (1) answering the query is the ultimate goal and can guarantee a non-negative reward and (2) asking a clarifying question does not have a high immediate reward and is risky but might eventually get a higher reward in the future from answering the query.

Now we explain the reinforcement learning algorithm. The goal of the risk control DQN learning is to train the DQN to predict the rewards $r_{ans}, r_{cq}$ of actions $a=\{\text{ANS},\text{CQ}\}$ given its input state $S$. And the training is to learn from all state-action-reward ($S$-$a$-$r_a$) tuples DQN has explored. Assume the action is to answer ($a=\text{ANS}$), the predicted reward of DQN is: \begin{equation}
y_{\text{pred}}(S)_{ans} = DQN(S)_{ans}
\end{equation}
The true reward $r$ is computed as the answer's reciprocal rank:

\begin{equation}
y_{\text{target}}(S)_{ans} = r_{ans} 
\end{equation}
In this case, since the action is to answer, the conversation will always end. Alternatively, if the action is to ask a clarifying question, and if the clarifying question is relevant, then:
\begin{equation}
y_{\text{target}}(S)_{cq} = r_{cq} + \sigma \cdot \max_{a^* = ans, cq} DQN(S^*)_{a^*}
\end{equation}
where $r_{cq}$ is the immediate reward for asking relevant questions, $\sigma$ is the discount factor for future reward, $S^*, a^*$ are the updated state and action based on current state $S$ and action $a$. The second term is essentially a discount factor times the higher predicted reward between answering the query and asking another clarifying question after updating the conversation state $S$ to $S^*$. 
If the action is to ask but the clarifying question is irrelevant, then:
\begin{equation}
y_{\text{target}}(S)_{cq} = p_{cq}
\end{equation}
where $p_{cq}$ is the penalty for irrelevant question.

Finally, the training goal is to minimize the average mean squared error loss between $y_{\text{pred}}(S)_{a}$ and target reward $y_{\text{target}}(S)_{a}$ over all the state-action-reward ($S$-$a$-$r_a$) tuples: 

\begin{equation}
L = \text{MSE}(y_{\text{target}}(S)_{a}, y_{\text{pred}}(S)_{a})
\end{equation}

Reinforcement learning is notorious for not being guaranteed to converge during training. To get a higher chance of convergence, we use the experience replay strategy. Experience replay is an often-used training strategy that revises previously learned samples and learns the new sample, making the training more likely to converge. In experience replay, we also increase the experience play times of action-reward pairs by asking clarifying questions. This can also be seen as data over-sampling. Empirically, we find that doing this can help the DQN to learn better about the risk of asking questions. We only train our DQN with actions that result in non-zero rewards. This includes correctly choosing relevant answers or clarifying questions, and wrongly choosing irrelevant clarifying questions. This increases the density of meaningful training samples during reinforcement learning.

\subsection{Inference}
During inference, our model will interact with the user until the conversation ends. Starting with the initial query from the user, the answer retrieval model and question retrieval model first retrieve the top relevant answers and clarifying questions as ranked lists. Then, the query, the conversation context, the top relevant answers, and questions are passed to the risk control model, which decides whether to answer the query or to ask a clarifying question. If the decision is to answer the query, then the top1 answer will be returned to the user. The conversation will end and we will evaluate the answer using the Reciprocal Rank. If the decision is to ask a clarifying question, then the top1 clarifying question will be returned to a user simulator. If the question is relevant, then the user simulator will return feedback to the question. We will update the context with additional information and iterate the entire process. If the question is irrelevant (is a negative sampled question), then the user simulator's patience will be consumed for once and the risk control model will try again with the irrelevant clarifying question removed from question candidates. Once the user's patience is used up, we will end the conversation. The inference result of one conversation can be in one of three cases: 1) query answered correctly. 2) query answered incorrectly. 3) user's patience is used up and the user leaves. We will describe how we evaluate inference results in the evaluation metrics section.

\section{Experiments}
We conduct simulation experiments as introduced in Section 3 to test whether our risk control conversational search agent can improve answer quality as well as improve users' search experience.

\subsection{Dataset}
We test our risk control conversational search agent on three datasets: MSDialog \cite{,msdialogrank,msdialogintent, msdialogintent2}, Ubuntu Dialog Corpus\cite{udcdataset}, and Opendialkg\cite{opendialkgdataset}. In this section. we will explain how we process each of them using some filtering options which we believe are necessary for conversational search scenarios. Then, we will give some example conversations from each dataset.

\subsubsection{\textbf{MSDialog}}
MSDialog dataset consists of question answering conversations from the Microsoft forum. We use the MSComplete subset, which contains the raw conversations. First, we need conversations between two users, namely one user and one agent, since two-participant conversations can best fit in the information need/answering/clarification scenario. In most conversations in the MSComplete set, more than two participants are involved, including the user who initiates the query and several Microsoft human agents. We assume that the agents all share the same goal of helping the user and will continue and supplement on each other, hence we simply consider all the different agents as one `agent' and reduce the number of participants down to two. After that, we find the conversations are not in turns, i.e., either the user or agent can talk several consecutive turns before the other responses. To make the conversation structurally cleaner, we merge all these consecutive turns into one by concatenating them. During this process, some conversation turns can grow too long because of multiple concatenations. We restrict that each turn can have only 512 tokens and the rest are truncated.

The above preprocessing makes all conversations in the dataset strictly with two participants and in turns. We also apply some filtering options to the dataset. Only conversations that meet the following criteria are kept in our final set: (1) The conversation needs to have a voted answer. In the MSDialog dataset, there is a binary label for each turn indicating whether the turn is voted as the final answer. We need the conversations to have voted answers to guarantee that conversations are complete. We also truncate all conversation turns after the voted answer turn since most of them are greetings so that we can focus on the process of information need/answering/clarification; (2) After truncating, the conversation needs to over 4 turns (each utterance is a turn) so that there is at least one clarifying question asked in the conversation; (3) After truncating, the conversation needs to be less than 10 turns for the speed of simulation experiments. Only a few outliers are left out by this filtering. 

After the above processing and filtering, we have a final set of 3,762 conversations which will be used for our simulation experiment. For the rest of the conversations that are missing a final answer or being too long, we set them aside and use them for re-ranker pre-training. 

Here we give an example of a conversation in the MSDialog dataset:

\begin{itemize}
    \item[\textbf{User:}] \textit{remsh.exe started to appear in my Task Manager and to have high CPU usage. [SEP] From two days I have remsh.exe in my Task Manager and it has high CPU usage. When I end process and restart Windows during rebooting OS makes some updates. I have Windows updates disabled as my PC is not compatible with the latest Win 10 updates/upgrades. For that reason reinstalling Windows is impossible option for me. I have CCleaner but older version than the compromised one. My antivir is Windows Defender and additionally I have Malwarebytes and Zemana. Google search says remsh.exe could be both Microsoft process and infection. Could you please help me?}
    
    \item[\textbf{Agent:}] \textit{Where exactly does remsh.exe live? Which folder?}
    
    \item[\textbf{User:}] \textit{I see folder C:$\backslash$Program Files$\backslash$rempl I ended the process so I can't open from Task Manager to see if this is the true file location. I noticed it starts alongside with Telemtry but this never happened before. Also Telemtry had two process opened this time.}

    \item[\textbf{Agent:}] \textit{This is not a native Microsoft process. Click the Startup tab in the Task Manager, then deactivate this program. When you right-click the folder C:$\backslash$Program Files$\backslash$rempl then you can see when you created it. This should give you a clue about its purpose.}
    
\end{itemize}

\subsubsection{\textbf{Ubuntu Dialog Corpus}}

Ubuntu Dialog Corpus (UDC) contains logs from Ubuntu-related chat rooms. Usually, the UDC dataset is used as response selection tasks where the task is to select the next response given the current dialog history. In this work, we use the UDC dataset in a slightly different task. The conversations in UDC have much more turns, but each turn is much shorter and sparser in meaning than the conversations in MSDialog. The reason for this difference is that the conversations in MSDialog are from Q\&A posts from the Microsoft forum, where both users and agents have enough time to write long queries and responses. However, conversations in UDC are from chat logs, where people prefer short sentences and respond to each other more frequently. Many turns are too short and not very meaningful. This makes the UDC dataset harder for simulation experiments as the information is very sparse. Another challenge in UDC dataset is that there is no identifier of the final answer as the `voted answer' field in MSDialog conversations. This brings several problems. The first is that we cannot separate the greetings from the main conversation. Also, we cannot guarantee that each conversation is complete with a final answer. Finally, there could be more than one conversation topic in the conversations.

For the UDC dataset, we apply the same processing and filtering options as in MSDialog on the raw conversations. The difference is that we do not merge participants in UDC conversations, but we only use conversations with exactly two participants. We still merge consecutive turns into one to make sure the conversations are in turns. Because UDC conversations are more concise than MSDialog, we filter out conversations where the minimum length among its turns is less than 5, i.e., we only keep a conversation when all of its turns have over 5 tokens. Because the UDC dataset is too large, we only randomly sample 10,000 conversations in our final set for the simulation experiment. We set the conversations filtered out aside and use them for pre-training re-rankers.

Here we give an example of a conversation in the UDC dataset:

\begin{itemize}
    \item[\textbf{User:}] \textit{Sigh, why do I always click the update button when it comes upOk, 12.04 have introduced an extreme blurryness to text, moving the mouse is really slow and everything is slow, it's like the graphics card is broken or something}
    
    \item[\textbf{Agent:}] \textit{what video chip do you use?}
    
    \item[\textbf{User:}] \textit{I'm running ubuntu in virtualbox, was perfect 11.10, now it's totally broken}

    \item[\textbf{Agent:}] \textit{tried changing text rendering options?}
    
    \item[\textbf{User:}] \textit{the text is blurry with green and red mixed in, it's like someone took subpixel rendering and made it backwards}
    
    \item[\textbf{Agent:}] \textit{if you search dash for the word: font does anything show?}
    
\end{itemize}

\subsubsection{\textbf{Opendialkg}}

The Opendialkg dataset contains conversations where a user asks an agent for a recommendation or opinion about a certain movie, music, books, etc. The Opendialkg dataset is originally used for conversational reasoning and knowledge graph entity prediction task. Apart from the raw conversation log data, the dataset also contains a knowledge graph, which is a key component of the original study. We do not use knowledge graph-related models in our experiments. Instead, we process the data as follows.

First, we extract the raw conversations and the knowledge graph entries from the dataset. The conversations are already strictly between two participants and in turns. The knowledge graph entries are relation tuples as "entity1, relation, entity2", for example, \textit{"Rose Bowl Stadium, Olympic games used in, 1984 Summer Olympic"}. We realize that the agent answers the user's need or raise clarifying question by looking up related information from the knowledge graph entries. To simulate this, for each user turn, we rank the relevance of knowledge graph entries as texts to the user's utterance text. We use BM25 for this ranking and we append the top 10 knowledge graph entries as texts to the end of user turn. It is not the most ideal way to use this dataset, but because the conversations in this dataset are of high quality compared to other datasets. So, we decide to give this a try.

An example of the knowledge enhanced conversation will be like this:
\begin{itemize}
    \item[\textbf{User:}] \textit{What movies has Baz Luhrmann directed? [SEP] Baz Luhrmann is-a Musician [SEP] Baz Luhrmann is-a Writer [SEP] The Great Gatsby (Award-Winning Work) directed\_by Baz Luhrmann [SEP] ...}
    
    \item[\textbf{Agent:}] \textit{Baz Luhrmann directed The Great Gatsby (Award-Winning Work). The Great Gatsby (Award-Winning Work) is starring Tobey Maguire. what else can i answer for you?}
    
    \item[\textbf{User:}] \textit{Was Tobey Maguire in anything else? [SEP] Tobey Maguire starred\_actors Spider-Man [SEP] Tobey Maguire starred\_actors Brothers [SEP] Tobey Maguire is-a Celebrity [SEP] ...}

    \item[\textbf{Agent:}] \textit{Tobey Maguire starred in The Cider House Rules. what else can i do for you today?}
    
    \item[\textbf{User:}] \textit{Wasn't he also in Spider-Man? [SEP] Abraham Benrubi starred\_actors The Man Who Wasn't There [SEP] Brooke Smith starred\_actors The Man Who Wasn't There [SEP] ...}
    
    \item[\textbf{Agent:}] \textit{Yes he was. what else can i answer for you today?}
    
    \item[\textbf{User:}] \textit{Thank you for your help! You are very helpful!}
    
    \item[\textbf{Agent:}] \textit{No problem!}
    
\end{itemize}

After this, we apply the same processing and filtering options as in MSDialog. We also truncate the greetings turns by using a greetings word list containing "welcome", "no problem", and so on. We set the conversations filtered out aside and use them for pre-training re-rankers. The Opendialkg dataset is originally not a response ranking dataset. Therefore, this dataset is the hardest dataset among the three, and the experiment is not as useful for reference as MSDialog and UDC.

We summarize all the three dataset statistics in Table ~\ref{datasetstat}, we will publish the processed dataset after the paper is published. 

\begin{table}[htbb]
\footnotesize
\caption{Dataset statistics}
\begin{center}
\begin{tabular}{|l|l|l|l|l|}
\hline
Dataset           & MSDialog (original) & MSDialog (ours) & UDC & Opendialkg \\
\hline
\# conversations & 35,000 & 3,762 & 10,000 & 10,000  \\
\hline
Max. turns        & 1700 & 10  & 10 & 10    \\
\hline
Min. turns        & 3 & 4  & 4 & 4     \\
\hline
Avg. turns        & 8.94 & 4.70  & 4.77 & 6.05 \\
\hline
Avg. \# clarifying questions & - & 1.7 & 1.77 & 2.05 \\ 
\hline
Avg. \# words per utterance  & 75.91 & 65.16 & 20.85 & 48.52  \\
\hline

\end{tabular}
\label{datasetstat}
\end{center}
\end{table}

\subsection{Baselines}
To show that our risk control conversational search agent can improve answer quality as well as improve user experience, we conduct experiments and compare it with multiple baselines:
\begin{enumerate}
    \item \textbf{Q0A}, a baseline that always answers the query asking no clarifying question. Hence, it always uses the initial query as only information for answer retrieval.
    
    \item \textbf{Q1A}, a baseline that always answers the query after asking exactly one clarifying question. It always asks a clarifying question in the 1st round, and answers in the second round. If it cannot ask any relevant clarifying question in the first round, the user will leave and it will stop there.
    
    \item \textbf{Q2A}, a baseline that always answers the query after asking exactly two clarifying questions, similar to Q1A.
    
    \item \textbf{CtxPred}, a risk-unaware baseline that first uses a binary classifier to decide whether to answer the query or ask a clarifying question given the context, then uses the corresponding retrieval model to generate a response. This is a common approach taken by previous works like \cite{xuasking2019, aliannejadi2020convai3}.
    
    \item \textbf{Oracle}, an ideal decision-making model, which is only for comparison. This oracle model can be seen as the upper bound of the model we are trying to train. This model never chooses the worse action, which is defined earlier in Section 3. Again, a worse action can be one of the two cases: (1) Asking an irrelevant question under any circumstance. (2) Answering with an irrelevant answer when there is a relevant question to ask and answer re-ranker's reciprocal rank is less than $1/t$, where $t$ is the user tolerance. This is theoretically the best risk control model for any fixed answer and question retrieval model.
\end{enumerate}

We test all these above baselines together with our risk control conversational search agent. During experiments, they share the same answer and question retrieval model and parameters, only the risk control decision-making parts are different.
\subsection{Technical Details}
In our experiment, we use the implementation of bi-encoder and poly-encoder from ParlAI\footnote{https://github.com/facebookresearch/ParlAI/tree/master/projects/polyencoder} with modifications to accommodate our experiments. We implement our risk control model and user simulator from scratch based on Pytorch. We split our datasets into 5 folds and use cross-validation to test significance. We run our main experiments on a single core of GeForce RTX 2080 Ti with 11GB memory. The pre-training of the bi-encoder and poly-encoder re-rankers are done on 4 of the above cores with a smaller batch size than the original settings of ParlAI. We test and tune our model hyper-parameters empirically. The number of negative samples $k=100$ for MSDialog and UDC experiments, and $k=20,50,100$ for Opendialkg experiments because of the difficulty in the Opendialkg dataset. We finally set the clarifying question reward $r_{cq}=0.11$ and penalty $p_{cq}=-0.89$, as this leads to the best MRR score. We set the future reward weight in reinforcement learning $\sigma = 0.89$ to make expected answer reward in future turns still in $[0,1]$. We train our model with a learning rate $lr=10^{-4}$, and regularization weight $\lambda=10^{-2}$ for the risk control DQN.

\section{Results and Analyses}

In this section, we will analyze our models and baselines with our risk-aware conversational search evaluation framework. The origin work \cite{wang2021controlling} only conducts experiments on the MSDialog dataset. Besides the MSDialog dataset, we also conduct experiments on the Ubuntu Dialog Corpus (UDC) and Opendialkg datasets. Besides showing that our risk control conversational search agent outperforms the baselines in most experiments, we also provide insights from the experiment results tables, figures, and analyses about why our model outperforms the baselines.

Our simulation experiments are done on three datasets, and on each dataset, we test two combinations of answer and question re-rankers with our risk control model. Then for each combination, we test different user simulator models with clarifying question patience $\rho=\infty, 2$ and tolerance $\tau=0,1,2$. These $\rho$ and $\tau$ choices are made based on the average number of clarifying questions of our dataset (1.7, 1.77, 2.05). There are in total six result tables, where Table ~\ref{msdialogpoly} and Table ~\ref{msdialogbi} are results of simulation on MSDialog dataset using poly-encoder and bi-encoder as answer/question re-ranker respectively. Table ~\ref{udcpoly} and ~\ref{udcbi} are results of simulation on UDC dataset using poly-encoder and bi-encoder as answer/question re-ranker respectively. Table ~\ref{openpoly} and Table ~\ref{openbi} are results of simulation on Opendialkg dataset using poly-encoder and bi-encoder as answer/question re-ranker respectively. In these tables, we show the results of both the baseline models and our models with two ablation study variations. `RCSQ' (Risk-aware Conversational Search agent with Q-learning) is our previous risk control model which uses both the textual and score features in risk control DQN. `RCSQ-S' (RCSQ minus score) is an ablation study model which removes the score features and only uses the textual features. `RCSQ-T' (RCSQ minus textual) is an ablation study model which removes the textual features and only uses the score features. In the tables, we abbreviate recall@1 in 100 candidates as R@1/100, Mean Reciprocal Rank as MRR, and decision error as Dec. err. We bold the best performance of each column and mark performances that significantly improve the best baseline model of the same column with $\dag$ and $\ddag$.

\subsection{MSDialog experiments}
The first simulation experiment (Table ~\ref{msdialogpoly}) is done on MSDialog using a poly-encoder re-ranker together with our risk control model. The first comparison we highlight is the comparison between Q0A and Q1A. We mentioned in data processing that we only keep conversations with more than 4 utterances, which means each conversation will have more than 2 conversation rounds. Therefore, there is always at least one clarifying question in each conversation, since we can be sure that only the last round is the answering round using the `voted answer' label of the MSDialog dataset as we mentioned earlier in the data processing. Naturally, the Q1A baseline which asks one clarifying question before giving an answer should perform better than Q0A, which directly retrieves the answer.
From the columns of $\tau=1$ and $\tau=2$ in the upper half of Table ~\ref{msdialogpoly}, we see that Q1A does improve over Q0A by having higher recall@1 ($0.4369>0.3994$ and $0.4602>0.3994$), MRR ($0.5277>0.5189$ and $0.5591>0.5189$) and lower decision error ($0.2841<0.3297$ and $0.2261<0.3089$). However, from the columns of $\tau=0$, we see that Q1A has lower recall@1 ($0.3794<0.3994$) and MRR score (0.4527<0.5189) and higher decision error rate (0.4089>0.3553). This result shows that there are indeed risks in asking clarifying questions. And the reason for this result is that the question retrieval model often cannot retrieve the relevant clarifying question, although the system thinks it is needed. The comparison between Q0A and Q1A shows that identifying clarifying question needs is not enough for improving answer quality and user experience. 

\begin{table}[t]

\caption{Comparison of all models and baselines on MSDialog dataset using poly-encoder as re-ranker. $\rho$ is the user patience for total clarifying questions, $\tau$ is the user tolerance for irrelevant clarifying questions. Numbers in bold mean the result is the best excluding oracle. $\ddag$ indicates $p < 0.01$ statistical significance over the best among baseline models.}
\scalebox{0.85}{
\begin{tabular}{l|l|l|l|l|l|l|l|l|l}
\hline\hline
\multirow{2}{*}{Users}  & \multicolumn{9}{c}{$\rho=\infty$} \\ 
\cline{2-10}  
& \multicolumn{3}{c|}{$\tau=0$} & \multicolumn{3}{c|}{$\tau=1$} & \multicolumn{3}{c}{$\tau=2$} \\ \hline
Models  & R@1/100     & MRR    & Dec. err    & R@1/100     & MRR    & Dec. err    & R@1/100     & MRR    & Dec. err  \\ \hline
Q0A     & 0.3994    & 0.5189    & 0.3553    & 0.3994    & 0.5189    & 0.3297    &  0.3994    & 0.5189    & 0.3089     \\ 
Q1A     & 0.3794    & 0.4527    & 0.4089    & 0.4367    & 0.5277   &  0.2841  & 0.4602    & 0.5591    & 0.2261   \\ 
Q2A     & 0.0722    & 0.0887    & 0.8844    & 0.0891   & 0.1102    &  0.8491  & 0.1030    & 0.1274    & 0.8211    \\ 
CtxPred & 0.3791    & 0.4523    & 0.3036   & 0.4366   & 0.5273   & 0.2033  & 0.4591    & 0.5587    & 0.1622            \\ \hline
RCSQ-S &  0.3919   & 0.4882    & 0.3236   & 0.4378 & 0.5284 & 0.2056  & 0.4611  & 0.5594    & 0.1631    \\ 
RCSQ-T &  0.4419  & 0.5428  & 0.2456   &  0.4519 &  0.5517  & 0.2350 &  0.4483 & 0.5505  & 0.2261  \\ 
RCSQ   & \textbf{0.4461}$^{\ddag}$    & \textbf{0.5435}$^{\ddag}$   & \textbf{0.2375}$^{\ddag}$ &  \textbf{0.4653}$^{\ddag}$ &  \textbf{0.5620}$^{\ddag}$ &   \textbf{0.1830}$^{\ddag}$ & \textbf{0.4788}$^{\ddag}$  & \textbf{0.5781}$^{\ddag}$  &  \textbf{0.1519}$^{\ddag}$           \\ \hline
Oracle  & 0.5033    & 0.6325    & 0  &   0.5472  &   0.6557     & 0    & 0.5589    & 0.6676  & 0   \\ \hline \hline

\multirow{2}{*}{Users}  & \multicolumn{9}{c}{$\rho=2$} \\ 
\cline{2-10}  
& \multicolumn{3}{c|}{$\tau=0$} & \multicolumn{3}{c|}{$\tau=1$} & \multicolumn{3}{c}{$\tau=2$} \\ \hline
Models  & R@1/100     & MRR    & Dec. err    & R@1/100     & MRR    & Dec. err    & R@1/100     & MRR    & Dec. err  \\ \hline
Q0A     & 0.4661    & 0.5910    & 0.2903    & 0.4661    & 0.5910    & 0.2694    &  0.4661    & 0.5910    & 0.2350     \\ 
Q1A     & 0.3836    & 0.4513    & 0.4378    & 0.4522    & 0.5369   &  0.3039  & 0.4836    & 0.5771    & 0.2378   \\ 
Q2A     & 0.0792   & 0.0937  & 0.8817    & 0.1039 & 0.1257 & 0.8347  & 0.1158    &0.1415    & 0.8056    \\ 
CtxPred & 0.3824    & 0.4508    & 0.4387   & 0.4510   & 0.5350   & 0.3068  & 0.4811    & 0.5737    & 0.2425            \\ \hline
RCSQ-S &  0.4661   & 0.5905    & 0.2911   & 0.4533 & 0.5560 & 0.2869  & 0.4758  & 0.5812    & 0.2419    \\ 
RCSQ-T &  \textbf{0.4774}$^{\ddag}$  & \textbf{0.5953}$^{\ddag}$  & \textbf{0.2542}$^{\ddag}$   &  0.4860 & 0.6030  & 0.2328 &  0.4884 & 0.6045  & \textbf{0.2035}$^{\ddag}$  \\ 
RCSQ   & 0.4758    & 0.5943   & 0.2678 &  \textbf{0.4947}$^{\ddag}$ &  \textbf{0.6045}$^{\ddag}$ &   \textbf{0.2292}$^{\ddag}$ & \textbf{0.5025}$^{\ddag}$  & \textbf{0.6064}$^{\ddag}$  &  0.2064           \\ \hline
Oracle  & 0.5522    & 0.6648    & 0  &   0.5736  &   0.6842     & 0    & 0.5850    & 0.6945  & 0   \\ \hline\hline

\end{tabular}
}
\label{msdialogpoly}
\end{table}

Then, we highlight the comparison between our model and the baselines. For all three user models with different tolerance, our model can outperform all the baselines (excluding the oracle) significantly. This shows that our risk control conversational search agent can retrieve the relevant answer to users more often as well as improve user experience. This is because our risk control model comprehensively evaluates the clarifying question need and the retrieved answer and clarifying question qualities before decision-making. Therefore, it has a lower decision error rate than all the baselines.

\begin{figure}[t]
  \centering
  \includegraphics[width=0.9\linewidth]{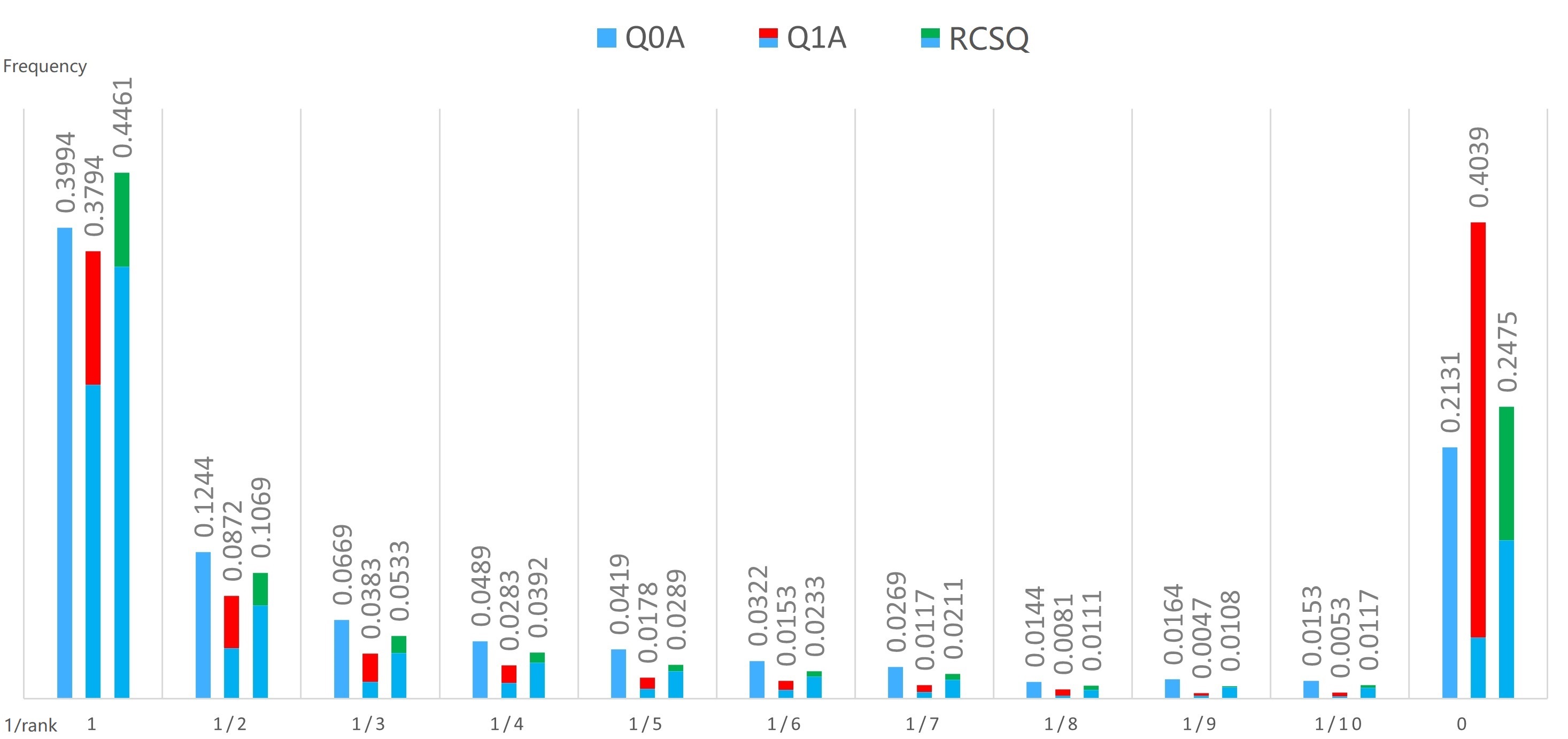}
  \caption{MRR Distributions of our model and baseline models in poly-re-ranker experiment with $\rho=\infty, \tau=0$ on MSDialog dataset. Blue proportion in Q1A and RCSQ represent the conversations that have the same results as Q0A, red/green proportion represent the conversations that have different result from Q0A, i.e., the improvements.}
  \Description{MRR Distributions for poly-re-ranker experiment and 0 tolerance on MSDialog dataset.}
  \label{polydistribution0}
\end{figure}

\begin{figure}[t]
  \centering
  \includegraphics[width=0.9\linewidth]{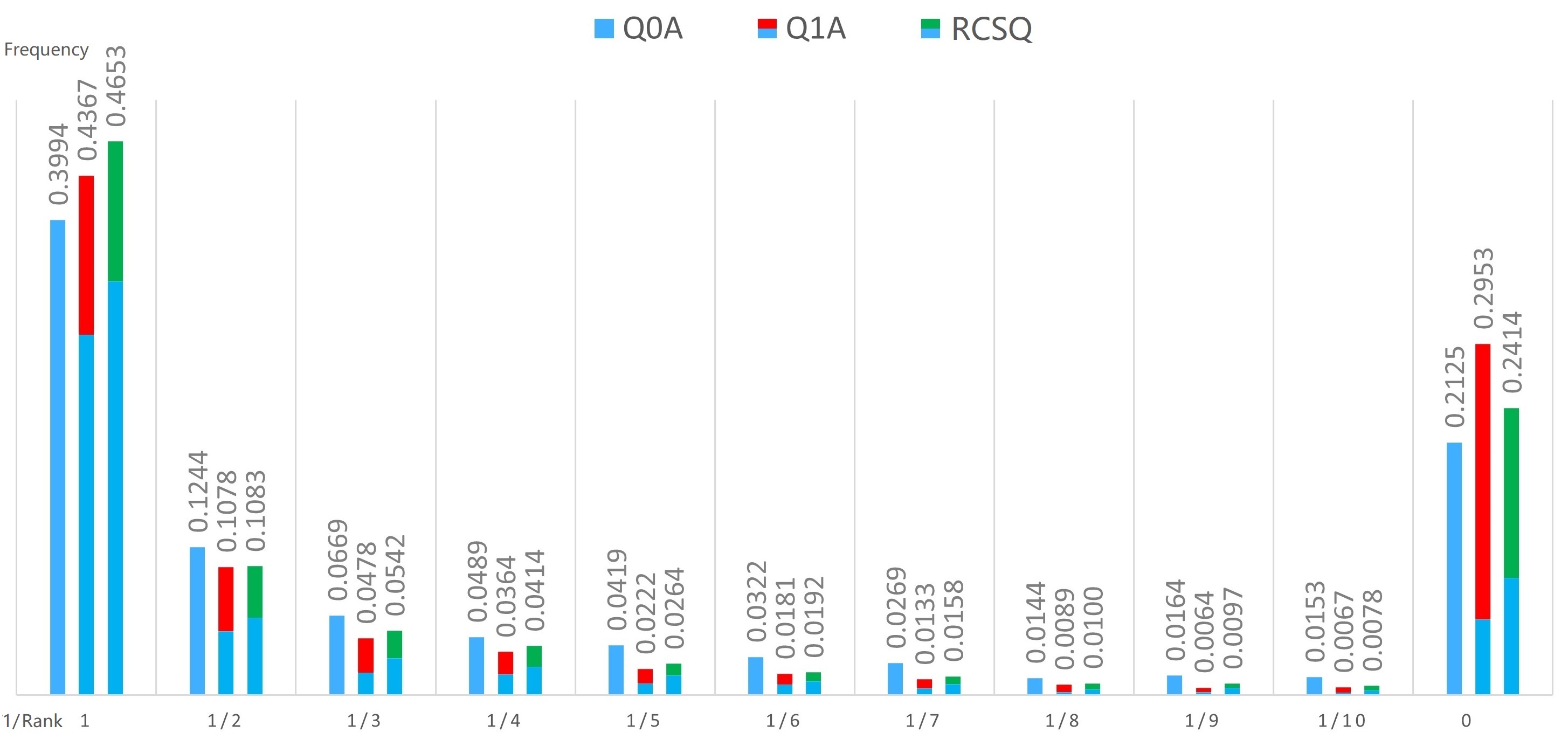}
  \caption{MRR Distributions of our model and baseline models in poly-re-ranker experiment with $\rho=\infty, \tau=1$ on MSDialog dataset. Blue proportion in Q1A and RCSQ represent the conversations that have the same results as Q0A, red/green proportion represent the conversations that have different result from Q0A, i.e., the improvements.}
  \Description{MRR Distributions for poly-re-ranker experiment and 1 tolerance on MSDialog dataset. }
  \label{polydistribution1}
\end{figure}

To better understand our model and the baseline models' behavior and why our model improves both the Q0A and Q1A baselines, we further study the distribution of MRR scores of our model, Q0A and Q1A baseline models on the test set when user tolerance is 0 and show the distribution in Figure ~\ref{polydistribution0}. From the figure, we can see the reason for each model's performance clearer. Now we explain the meaning of the coloring. The numbers over the bars are the total frequencies, e.g., in Fig 3, the leftmost bar's frequency 0.3994 means that the answer returned by Q0A gets MRR=1 in 39.94\% of the test conversations. The second bar's frequency of 0.3794 means that the answer returned by Q1A (in its second turn) gets MRR=1 in 37.94\% of the test conversations. Among the 37.94\%, there are some conversations where Q0A also gets MRR=1, and they are colored in blue. The rest of the 37.94\% are the `new' conversations that only Q1A gets MRR=1 while Q0A cannot get MRR=1 (Q0A may get any MRR lower than 1 in these conversations), and they are colored in red. Similarly, in RCSQ, the blue part is conversations that Q0A also gets MRR=1, and the green part is where only RCSQ gets MRR=1 and Q0A not (now the turn of when the answer is returned is non-deterministic and is decided by RCSQ).
    
The red and green part of the bars can be seen as the benefit of asking clarifying questions. However, asking clarifying questions also bears the risk of returning irrelevant questions and decreasing MRR. This can be seen in the figure by that the blue parts of Q1A and RCSQ are always shorter than that of Q0A. These `lost' parts are when asking clarifying questions harm answer quality and decrease MRR. For example, comparing the leftmost blue bar of Q0A and the blue part of the 2nd leftmost bar of Q1A, we see that Q1A loses a sizeable chunk of conversations by asking irrelevant questions. These `lost' chunk of conversations are where the 1st turn answer is already good but asking clarifying questions is bad (and this lost chuck mainly goes to the red part of Q1A in the MRR=0 cell). This exactly explains why Q1A does not outperform Q0A when $\tau=0$. If we add the 3rd left-most bar of RCSQ into comparison, we can see why RCSQ is much better than Q1A. We can see a drop of the blue bar in RCSQ, but it is far less than the drop on Q1A's bar. Reducing this part is the actual meaning of controlling the risk of conversational search and is what we believe is missing from other clarifying question studies about conversational search.
    
When we stack the blue and red/green bars together, we combine the benefit and harm of asking clarifying questions together. Now the total heights of these bars measure the overall effect of each system on answer quality. We see that the green bar gaining in RCSQ is more than the blue bar loss. This means that RCSQ can maximize the benefit of clarifying questions while controlling their harm. Thus, it outperforms both Q0A and Q1A.

Here is an example of how our model improves the baselines by choosing the better action between answering and asking questions.

\begin{itemize}
    \item[\textbf{Case 1}]
    \item[Round 1]
    \item[User:] (Initial Query) \textit{Div/0 Error In Access. [SEP] I have a form that displays a Div/0 error when one the required fields is left at 0. The calculation is done in the background query. Is there a way to have the form display a blank or a N/A when a Div/0 error occurrs?}
    
    \item[Top Ans:] \textit{This VLOOKUP should work then (assuming Sheet 2 is actually called "Sheet 2" with a space before the 2. In cell B2, enter this: =IFERROR(VLOOKUP(A2,'Sheet 2'!A:D,2,FALSE),"") In C2: =IFERROR(VLOOKUP(A2,'Sheet 2'!A:D,3,FALSE),"") And in D2: =IFERROR(VLOOKUP(A2,'Sheet 2'!A:D,3,FALSE),"") The IFERROR function shows a blank in each cell if nothing is picked in the dropdowns in column A. You'll just need to copy the formulas down the columns.}
    
    \item[Top CQ:] \textit{What is the expression in the query? You should be able to amend it to incorporate an IIF function call so that the division by zero error is avoided. Or you can write a little function which does the same.}

    \item[\textbf{Analysis}:]
    At this stage, Q0A will choose to answer with the top retrieved answer which is not relevant to the query because the answer retriever lacks sufficient context information to retrieve the best answer. In this case, the correct answer is ranked 8th by answer re-ranker, thus Q0A gets $\text{R@1/100}=0$, MRR=0.125, and Dec.err.=1. 
    
    Q1A, Q2A, and RCSQ will respond the user with the top retrieved question which is a follow-up question in the original conversation. Hence user will reply the question with more information, and the conversation will continue.
    
    \item[Round 2]
    \item[User:] (Reply) \textit{The expression in the query field is Task Productivity Rate: [Total Quantity]/[Total Hours]}
    
    \item[Top Ans:] \textit{The following should return a zero if Total Hours is zero: IIF([Total Hours]>0,[Total Quantity]/[Total Hours], 0) or to return 'N/A': IIF([Total Hours]>0,[Total Quantity]/[Total Hours],"N/A") The former might be better if you need to do any further operations on the value e.g. summation over a set of rows in a report. PS: The above should work in a query, but not in code.}
    
    \item[Top CQ:] \textit{Dennis here is an example for A31:F64 as as 85\% A168:J202 as 40\% All other areas were 100 Sub PrintPreview01() 'June 03, 2015 Dim r As Range, v As Variant, r1 As Range Dim ws As Worksheet Set ws = ActiveSheet v = Array(100, 85, 100, 100, 100, 40, 100, 100) i = 0 Set r1 = ws.Range("A1:I30,A31:F64,A65:F88,A89:H135,A136:G167,A168:J202,A203:I239,A240: G264") For Each r In r1.Areas With ws.PageSetup .PrintArea = r.Address .Zoom = v(i) End With ws.PrintPreview i = i + 1 'ws.PrintOut Next End Sub}
    
     \item[\textbf{Analysis}:]
    At this stage, RCSQ chooses to answer with the top retrieved answer which is the correct answer to the query. So does Q1A. They both get $\text{R@1/100}=1$, MRR=1, and Dec.err.=0. Q2A chooses to ask the top retrieved question, but the question is irrelevant. Thus user will end the search session. Q2A gets $\text{R@1/100}=0$, MRR=0, and Dec.err.=0.5, because Q2A is correct on asking the first question but wrong on asking the second.
\end{itemize}

When we increase user tolerance $\tau$ to 1, the new distribution is as shown in Figure~\ref{polydistribution1}. From the figure, we see that the red bar in group 0 is much lower than the red bar in Figure~\ref{polydistribution0} and generally, Q1A's performance is better than Q0A since it can ask fewer irrelevant clarifying questions when the user gives it one more chance. We can also see that the green bar in group 1 in Figure~\ref{polydistribution1} is higher than the green bar in group 1 in Figure~\ref{polydistribution0}. This tells us that when the user becomes more tolerant, there are more conversations where our model can improve an irrelevant retrieved answer to a relevant one by asking clarifying questions. From this figure, we see that our model is still better than both of the baselines. The reasons are the same as in Figure~\ref{polydistribution1}.

We then decrease user patience $\rho$ to 2 let it interact with our model and the baseline models. This experiment is shown in the lower half of Table~\ref{msdialogpoly}. This set of experiments uses different random split data from the upper half of the table because we lost the original random split data. However, both experiments show the same trend and conclusions that RCSQ improves baseline models when interacting with different user tolerance settings on all evaluation metrics. This further shows that the RCSQ model's advantage is independent of user types.

Our risk control DQN has a long feature list consisting of both textual features and score features. The query, context, answer, and questions textual features are first encoded by DQN and then concatenated with score features to compute the action rewards. A question in DQN is that the score features from poly-encoder have already contained context-answer and context-question relevance. Are the textual features or the score features really necessary to be included twice? To answer this question, we do ablation studies on the feature list and test two variations of our risk control model: textual features $(q, h, cq^1,..., cq^k, a^1,...,a^k)$ only and score features $(s_{cq}^{1:k}, s_{ans}^{1:k})$ only. The two ablation study models are noted as RCSQ-S and RCSQ-T in Table~\ref{msdialogpoly}. From the table, we can see that all three variations of our model can outperform all the baselines in almost all the experiments except a few cells. Our previous model, which uses both textual and score features, in most cases outperforms the other variations. This answers the previous question of whether to include both features, and the answer is yes. Also, we find that when we increase user tolerance, the significance of textual features increases while the significance of score features declines in general.

Some observations in the tables are interesting or hard to understand at the first glance. The most important observation is that as user tolerance increases, the recall@1 improvement of our model over the best of baselines decrease ($0.0467>0.0287>0.0186$). This result is in our expectation. Because when users have more tolerance for irrelevant clarifying questions, the risk of asking a clarifying question will decrease. Then having a risk control model will be less useful. In the extreme case where users have an infinite tolerance for clarifying questions, there will be no risk in asking a clarifying question. The best agent strategy will be to ask clarifying questions until a relevant question is asked and then use users' feedback to retrieve better answers. In this extreme case, Q1A, Q2A, Oracle, and our model will all converge to the same performance. 

The second observation is that the Ctxpred baseline's performance is almost the same as the Q1A baseline. The reason is that the Ctxpred baseline is trained to predict whether it is necessary to ask clarifying questions only given the conversation context. All of our conversations involve at least one clarifying question, thus a well-trained Ctxpred model will always ask at least one clarifying question and potentially more, which leads to that the decision-making of Ctxpred the same as Q1A in the first conversation round. Thus, their behavior is very similar to each other. This also shows that the risk-unaware Ctxpred baseline is not the best solution to leveraging asking clarifying questions in conversational search. We assume that the Ctxpred baseline will work better in a dataset where asking at least one clarifying question is not always necessary. 

The third observation is that Q0A's decision error decreases as user tolerance increases. This one seems hard to understand at the first glance because Q0A's decision is fixed and independent of user tolerance. The actual reason is in the definition of decision error that when tolerance $t$ increases, the answer reciprocal rank threshold $1/t$ decreases. Thus, when Q0A retrieves irrelevant questions, they are considered not as worse decisions as asking irrelevant questions by the definition. The last observation is that Q2A performs significantly worse than Q1A. There are two reasons for it. The first reason is that the average number of turns in the MSDialog dataset is 4.7, which means most conversation only has only 2 rounds, i.e., only one clarifying question is asked. Q2A is guaranteed to be wrong in these conversations. The other reason is that Q2A needs to retrieve the first question correctly and then retrieve the second question correctly, and finally, the answer correctly. Q1A only needs to be correct twice on the first question and the answer, while Q0A needs only to be correct once. The difficulties in Q0A, Q1A, and Q2A grow exponentially. 

\begin{table}[t]
\caption{Comparison of all models and baselines on MSDialog dataset using bi-encoder as re-ranker. $\rho$ is the user patience for total clarifying questions, $\tau$ is the user tolerance for irrelevant clarifying questions. Numbers in bold mean the result is the best excluding oracle. $\ddag$ indicates $p < 0.01$ statistical significance over the best among baseline models.}
\scalebox{0.85}{
\begin{tabular}{l|l|l|l|l|l|l|l|l|l}
\hline\hline
\multirow{2}{*}{Users}  & \multicolumn{9}{c}{$\rho=\infty$} \\ 
\cline{2-10}  
& \multicolumn{3}{c|}{$\tau=0$} & \multicolumn{3}{c|}{$\tau=1$} & \multicolumn{3}{c}{$\tau=2$} \\ \hline
Models  & R@1/100     & MRR    & Dec. err    & R@1/100     & MRR    & Dec. err    & R@1/100     & MRR    & Dec. err  \\ \hline
Q0A     & 0.3653    & 0.4918    & 0.3444    & 0.3653    & 0.4918    &  0.3253   &  0.3653    & 0.4918    & 0.3061     \\ 
Q1A     & 0.3272    & 0.3994    & 0.4733    & 0.3914 & 0.4805  & 0.3494 & 0.4192 & 0.5180 & 0.2742   \\ 
Q2A     & 0.0639  & 0.0792  & 0.8922    & 0.0886   & 0.1107   &  0.8444  & 0.1019  & 0.1275  & 0.8181    \\ 
CtxPred & 0.3250  & 0.3968    & 0.4764   & 0.3869 & 0.4757 & 0.3558  & 0.3989  & 0.4928  & 0.3100 \\ \hline
RCSQ-S    & 0.3661   & 0.4924   & 0.3439 & 0.3881 & 0.4808 & 0.2894 & 0.4200 & 0.5192 & 0.2739  \\ 
RCSQ-T    & 0.3908   & 0.5086   & 0.2964 & 0.3917 & 0.5119 & 0.2739 & 0.4047 & 0.5228 & 0.2586  \\
RCSQ   & \textbf{0.4036}$^{\ddag}$  & \textbf{0.5165}$^{\ddag}$ & \textbf{0.2838}$^{\ddag}$ & \textbf{0.4253}$^{\ddag}$  & \textbf{0.5325}$^{\ddag}$  & \textbf{0.2625}$^{\ddag}$  &  \textbf{0.4314}$^{\ddag}$ &  \textbf{0.5416}$^{\ddag}$  &  \textbf{0.2306}$^{\ddag}$ \\ \hline
Oracle  & 0.4825    & 0.5984   & 0  & 0.5039  & 0.6244  & 0   & 0.5228  & 0.6368  & 0   \\ \hline\hline
\multirow{2}{*}{Users}  & \multicolumn{9}{c}{$\rho=2$} \\ 
\cline{2-10}  
& \multicolumn{3}{c|}{$\tau=0$} & \multicolumn{3}{c|}{$\tau=1$} & \multicolumn{3}{c}{$\tau=2$} \\ \hline
Models  & R@1/100     & MRR    & Dec. err    & R@1/100     & MRR    & Dec. err    & R@1/100     & MRR    & Dec. err  \\ \hline
Q0A     & 0.3653    & 0.4918    & 0.3444    & 0.3653    & 0.4918    &  0.3353   &  0.3653    & 0.4918    & 0.3061    \\ 
Q1A     & 0.3272    & 0.3994    & 0.4733    & 0.3914 & 0.4805  & 0.3494 & 0.4192 & 0.5180 & 0.2742   \\ 
Q2A     &  0.0639  & 0.0792  & 0.8922    & 0.0886   & 0.1107   &  0.8444  & 0.1019  & 0.1275  & 0.8181    \\ 
CtxPred & 0.3250  & 0.3968    & 0.4764   & 0.3869 & 0.4757 & 0.3558  & 0.3989  & 0.4928  & 0.3100            \\ \hline
RCSQ-S &  0.3636   & 0.4876    & 0.3511   & 0.3847 & 0.4800 & 0.3558  & 0.4175  & 0.5154    & 0.2786    \\ 
RCSQ-T &  0.3787  & 0.5005  & 0.3197 &  0.3910 & 0.5120  & 0.2897 &  0.3946 & 0.5163  & 0.2637  \\ 
RCSQ   & \textbf{0.3961}$^{\ddag}$    & \textbf{0.5021}$^{\ddag}$   & \textbf{0.3106}$^{\ddag}$ &  \textbf{0.4217}$^{\ddag}$ &  \textbf{0.5247}$^{\ddag}$ &   \textbf{0.2739}$^{\ddag}$ & \textbf{0.4283}$^{\ddag}$  & \textbf{0.5351}$^{\ddag}$  &  \textbf{0.2417}$^{\ddag}$           \\ \hline
Oracle  & 0.4825    & 0.5984   & 0  & 0.5039  & 0.6244  & 0   & 0.5228  & 0.6368  & 0   \\ \hline\hline
\end{tabular}
}
\label{msdialogbi}
\end{table}

Our second experiment is done on MSDialog using the bi-encoder as the answer and question retrieval model. Bi-encoder does not have a context-candidate attention mechanism, thus its performance is usually worse than poly-encoder in re-ranking tasks. In Table~\ref{msdialogbi}, we can see that the bi-encoder's performance is indeed worse than the poly-encoder in Table~\ref{msdialogpoly}. However, we see that it still holds that our model is better than the baseline models in all the columns. This shows that the effect of our risk control model does not depend on the answer/question retrieval model, as it should not, because it only aims to improve the performances based on these retrieval models. The observations from MSDialog poly-encoder experiment in Table~\ref{msdialogpoly} also can be seen in this table. These observations further strengthen our assumptions and explanations in the MSDialog poly-encoder experiment.

Among the ablation study models, our model, which uses both the textual and score features, is still the best of all variations in most experiments. We also see the same trend in `RCSQ-S' and `RCSQ-T' as in the poly-encoder experiment that as user tolerance increases, the textual features become more useful and the score features become less useful.

From our experiments on the MSDialog dataset, we show that our risk-aware conversational search agent which uses the risk control model along with the existing answer and question retrieval models is a better strategy for conversational search. And using both the textual features and the score features as inputs for the risk control model is the best configuration.

\subsection{UDC experiments}
The origin work \cite{wang2021controlling} only conducts experiments on the MSDialog dataset. Besides the MSDialog dataset, we also conduct experiments on the Ubuntu Dialog Corpus (UDC) dataset. In Table~\ref{udcpoly} and Table~\ref{udcbi}, we show results of experiments on the UDC dataset using poly-encoder and bi-encoder, respectively. Since our simulation experiment is a different task from the original response selection task on UDC, and we do not use the whole UDC dataset to train our re-rankers, we remind that the results in Table~\ref{udcpoly} and Table~\ref{udcbi} should not be directly compared to the state-of-the-art models' performances \cite{henderson2019convertonUDC, polyencoder,chen2019sequentialonUDC} on UDC dataset.

\begin{table}[t]
\caption{Comparison of all models and baselines on sampled UDC dataset using poly-encoder as re-ranker. $\rho$ is the user patience for total clarifying questions, $\tau$ is the user tolerance for irrelevant clarifying questions. Numbers in bold mean the result is the best excluding oracle. $\dag$ and $\ddag$ means $p < 0.1$ and $p < 0.05$ statistical significance over the best among baseline models.}
\scalebox{0.85}{
\begin{tabular}{l|l|l|l|l|l|l|l|l|l}
\hline\hline
\multirow{2}{*}{Users}  & \multicolumn{9}{c}{$\rho=\infty$} \\ 
\cline{2-10}  
& \multicolumn{3}{c|}{$\tau=0$} & \multicolumn{3}{c|}{$\tau=1$} & \multicolumn{3}{c}{$\tau=2$} \\ \hline
Models  & R@1/100     & MRR    & Dec. err    & R@1/100     & MRR    & Dec. err    & R@1/100     & MRR    & Dec. err  \\ \hline
Q0A     & 0.1580    & 0.2381    & 0.2870    & 0.1580    & 0.2381    & 0.3315    &  0.1580    & 0.2381    & 0.3610     \\ 
Q1A     & 0.1200    & 0.1630    & 0.6315    & 0.1520    & 0.2051   &  0.5330  & 0.1700    & 0.2305    & 0.4640   \\ 
Q2A     & 0.0230   & 0.0284  & 0.9525  & 0.0320  & 0.0417  & 0.9220  &  0.0430 & 0.0563  & 0.8940     \\ 
CtxPred & 0.1195  & 0.1625  & 0.6325  & 0.1510   & 0.2037   & 0.5360 & 0.1680    & 0.2278 & 0.4705          \\ \hline
RCSQ-S    & 0.1585   & 0.2382   & 0.2895 & 0.1545  & 0.2323 & 0.3575 & 0.1700   & 0.2307  & 0.4645    \\ 
RCSQ-T    &  0.1660   & \textbf{0.2452}   & 0.2590  & 0.1700  & 0.2483 & 0.2895  & 0.1710 & 0.2502  & 0.3220   \\
RCSQ    & \textbf{0.1675}$^{\ddag}$   & 0.2449$^{\ddag}$   & \textbf{0.2375}$^{\ddag}$  &  \textbf{0.1745}$^{\ddag}$ &  \textbf{0.2501}$^{\ddag}$  &  \textbf{0.2705}$^{\ddag}$ & \textbf{0.1810}$^{\ddag}$  & \textbf{0.2578}$^{\ddag}$  & \textbf{0.2905}$^{\ddag}$  \\ \hline
Oracle  & 0.2100    & 0.2985    & 0  &   0.2260  &   0.3163     & 0    & 0.2370    & 0.3301  & 0   \\ 
\hline\hline\multirow{2}{*}{Users}  & \multicolumn{9}{c}{$\rho=2$} \\ 
\cline{2-10}  
& \multicolumn{3}{c|}{$\tau=0$} & \multicolumn{3}{c|}{$\tau=1$} & \multicolumn{3}{c}{$\tau=2$} \\ \hline
Models  & R@1/100     & MRR    & Dec. err    & R@1/100     & MRR    & Dec. err    & R@1/100     & MRR    & Dec. err  \\ \hline
Q0A     & 0.1580    & 0.2381    & 0.2870    & 0.1580    & 0.2381    & 0.3315    &  0.1580    & 0.2381    & 0.3610     \\ 
Q1A     & 0.1200    & 0.1630    & 0.6315    & 0.1520    & 0.2051   &  0.5800  & 0.1700    & 0.2305    & 0.5215   \\ 
Q2A     & 0.0230   & 0.0284  & 0.9525  & 0.0320  & 0.0417  & 0.9220  &  0.0430 & 0.0563  & 0.8940     \\ 
CtxPred & 0.1165  & 0.1581  & 0.6755  & 0.1505   & 0.2033   & 0.5830 & 0.1690    & 0.2294 & 0.5235       \\ \hline
RCSQ-S    & 0.1570   & 0.2369   & 0.2890 & 0.1575  & 0.2369 & 0.3340 & 0.1575 & 0.2379 & 0.3610 \\ 
RCSQ-T    &  0.1600   & 0.2392 & 0.2795   & 0.1660  & \textbf{0.2467}$^{\ddag}$  & \textbf{0.3020}$^{\ddag}$& 0.1691 & 0.2487  & \textbf{0.3307}$^{\ddag}$   \\
RCSQ    &  \textbf{0.1650}$^{\ddag}$   & \textbf{0.2451}$^{\ddag}$   & \textbf{0.2655}$^{\ddag}$  &  \textbf{0.1725}$^{\ddag}$ &  0.2381 &  0.4340 & \textbf{0.1890}$^{\ddag}$   & \textbf{0.2599}$^{\ddag}$  & 0.3750\\ \hline
Oracle  & 0.2155    & 0.3029    & 0  &   0.2345  &   0.3232     & 0    & 0.2490   & 0.3406 & 0   \\ 

\hline\hline
\end{tabular}
}
\label{udcpoly}
\end{table}

\begin{table}[t]
\caption{Comparison of all models and baselines on sampled UDC dataset using bi-encoder as re-ranker. $\rho$ is the user patience for total clarifying questions, $\tau$ is the user tolerance for irrelevant clarifying questions. Numbers in bold mean the result is the best excluding oracle. $\dag$ and $\ddag$ means $p < 0.1$ and $p < 0.05$ statistical significance over the best among baseline models.}
\scalebox{0.85}{
\begin{tabular}{l|l|l|l|l|l|l|l|l|l}
\hline\hline
\multirow{2}{*}{Users}  & \multicolumn{9}{c}{$\rho=\infty$} \\ 
\cline{2-10}  
& \multicolumn{3}{c|}{$\tau=0$} & \multicolumn{3}{c|}{$\tau=1$} & \multicolumn{3}{c}{$\tau=2$} \\ \hline
Models  & R@1/100     & MRR    & Dec. err    & R@1/100     & MRR    & Dec. err    & R@1/100     & MRR    & Dec. err  \\ \hline
Q0A     & 0.1335    & 0.2138    & 0.3340    & 0.1335    & 0.2138    & 0.3990     & 0.1335    & 0.2138    & 0.4315  \\ 
Q1A     & 0.0995    & 0.1478    & 0.6295    & 0.1215    &  0.1844  & 0.5285    & 0.1385  & 0.2097   & 0.4715   \\ 
Q2A     & 0.0185    & 0.0264    & 0.9470    & 0.0260   & 0.0368    & 0.9160    &   0.0355  &  0.0506   & 0.8825     \\ 
CtxPred &0.0995    & 0.1478    & 0.6295  & 0.1195   & 0.1807   & 0.5385  & 0.1385  & 0.2093   & 0.4735  \\ \hline
RCSQ-S    & 0.1345    & 0.2145   & 0.3335 & 0.1335  & 0.2142 & 0.3980 & 0.1385  & 0.2094 &  0.4730      \\ 
RCSQ-T    & 0.1415    & 0.2197   & 0.3055 & 0.1415  & 0.2212 & 0.3745  & 0.1470  & 0.2236  &  0.3950      \\
RCSQ    & \textbf{0.1420}$^{\ddag}$     & \textbf{0.2201}$^{\ddag}$   & \textbf{0.2995}$^{\ddag}$  &  \textbf{0.1465}$^{\ddag}$ & \textbf{0.2258}$^{\ddag}$ & \textbf{0.3460}$^{\ddag}$  & \textbf{0.1525}$^{\ddag}$   & \textbf{0.2303}$^{\ddag}$ &  \textbf{0.3690}$^{\ddag}$     \\ \hline
Oracle  & 0.1840    & 0.2779    & 0  & 0.1995    & 0.2975       & 0    &   0.2135  & 0.3133  & 0   \\ 
\hline\hline\multirow{2}{*}{Users}  & \multicolumn{9}{c}{$\rho=2$} \\ 
\cline{2-10}  
& \multicolumn{3}{c|}{$\tau=0$} & \multicolumn{3}{c|}{$\tau=1$} & \multicolumn{3}{c}{$\tau=2$} \\ \hline
Models  & R@1/100     & MRR    & Dec. err    & R@1/100     & MRR    & Dec. err    & R@1/100     & MRR    & Dec. err  \\ \hline
Q0A     & 0.1335    & 0.2138    & 0.3340    & 0.1335    & 0.2138    & 0.3990     & 0.1335    & 0.2138    & 0.4315   \\ 
Q1A     & 0.0995    & 0.1478    & 0.6295    & 0.1215    & 0.1844   &  0.5285  & 0.1385    & 0.2097    & 0.4715   \\ 
Q2A     & 0.0185  & 0.0264  & 0.9470  & 0.0260  & 0.0368  & 0.9160  &  0.0355 & 0.0506  & 0.8825     \\ 
CtxPred & 0.1165  & 0.1581  & 0.6755  & 0.1190   & 0.1810   & 0.5370 & 0.1385    & 0.2093 & 0.4730       \\ \hline
RCSQ-S    & 0.1330   & 0.2132   & 0.3340 & 0.1360  & 0.2081 & 0.4415 & 0.1390 & 0.2099 & 0.4715 \\ 
RCSQ-T    &  0.1405   & 0.2187 & 0.3110   & \textbf{0.1425}$^{\ddag}$  & \textbf{0.2209}$^{\ddag}$  & 0.3665& 0.1421 & 0.2199  & 0.3907   \\
RCSQ    &  \textbf{0.1420}$^{\ddag}$   & \textbf{0.2190}$^{\ddag}$   & \textbf{0.2970}$^{\ddag}$  &  0.1410 &  0.2189 &  \textbf{0.3625}$^{\ddag}$ & \textbf{0.1490}$^{\ddag}$   & \textbf{0.2273}$^{\ddag}$  & \textbf{0.3760}$^{\ddag}$\\ \hline
Oracle  & 0.1840    & 0.2779    & 0  &   0.1995  &   0.2975     & 0    & 0.2135   & 0.3133 & 0   \\ 
\hline\hline
\end{tabular}
}
\label{udcbi}
\end{table}

As discussed in the dataset section, there are many challenges in the UDC dataset which make it less ideal and harder than the MSDialog dataset. When we compare the baseline performances on UDC with their performances on MSDialog, we see that all the baselines have a harder time on UDC datasets. For example, the Q0A baseline retrieves correct answers from 100 candidate answers for $39.94\%$ of the conversations in MSDialog, but only $15.80\%$ conversations in UDC. Similar comparisons can be found for Q1A, Q2A, and the Oracle model. This shows that the UDC dataset is indeed a harder dataset for simulation experiments.

Despite the UDC conversations being harder and the retrieval models being less successful on the UDC dataset, we still see that the RCSQ model significantly outperforms the baseline models when interacting with different user types. This shows that our risk control model is not only independent of the retrieval model structure but also independent of the efficiency of the retrieval models. This is because no matter how optimized the answer and question retrieval models are, there will always be conversations where either the answer or question retrieval model retrieves significantly better responses than the other. Our risk control model is trained to evaluate their retrieval results qualities and choose the better retrieval result. Thus, our risk control model performance is better than all the baselines without this decision-making flexibility.

\begin{figure}[ht]
  \centering
  \includegraphics[width=0.9\linewidth]{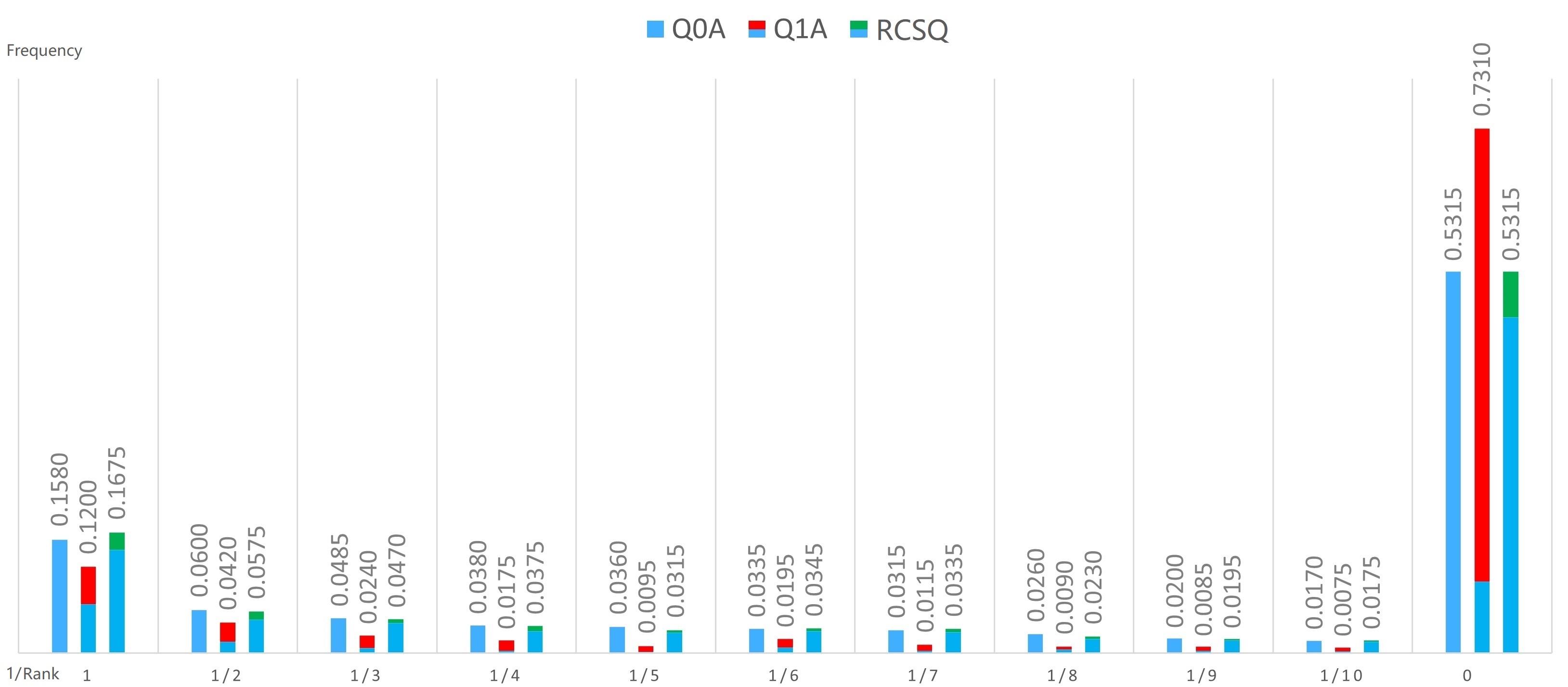}
  \caption{MRR Distributions of our model and baseline models in poly-re-ranker experiment with user $\rho=\infty, \tau=0$ on UDC dataset. Blue proportion in Q1A and RCSQ represent the conversations that have the same results as Q0A, red/green proportion represent the conversations that have different result from Q0A, i.e., the improvements.}
  \Description{MRR Distributions for poly-re-ranker experiment and 0 tolerance on UDC dataset.}
 \label{udcdistribution0}
\end{figure}

\begin{figure}[ht]
  \centering
  \includegraphics[width=0.9\linewidth]{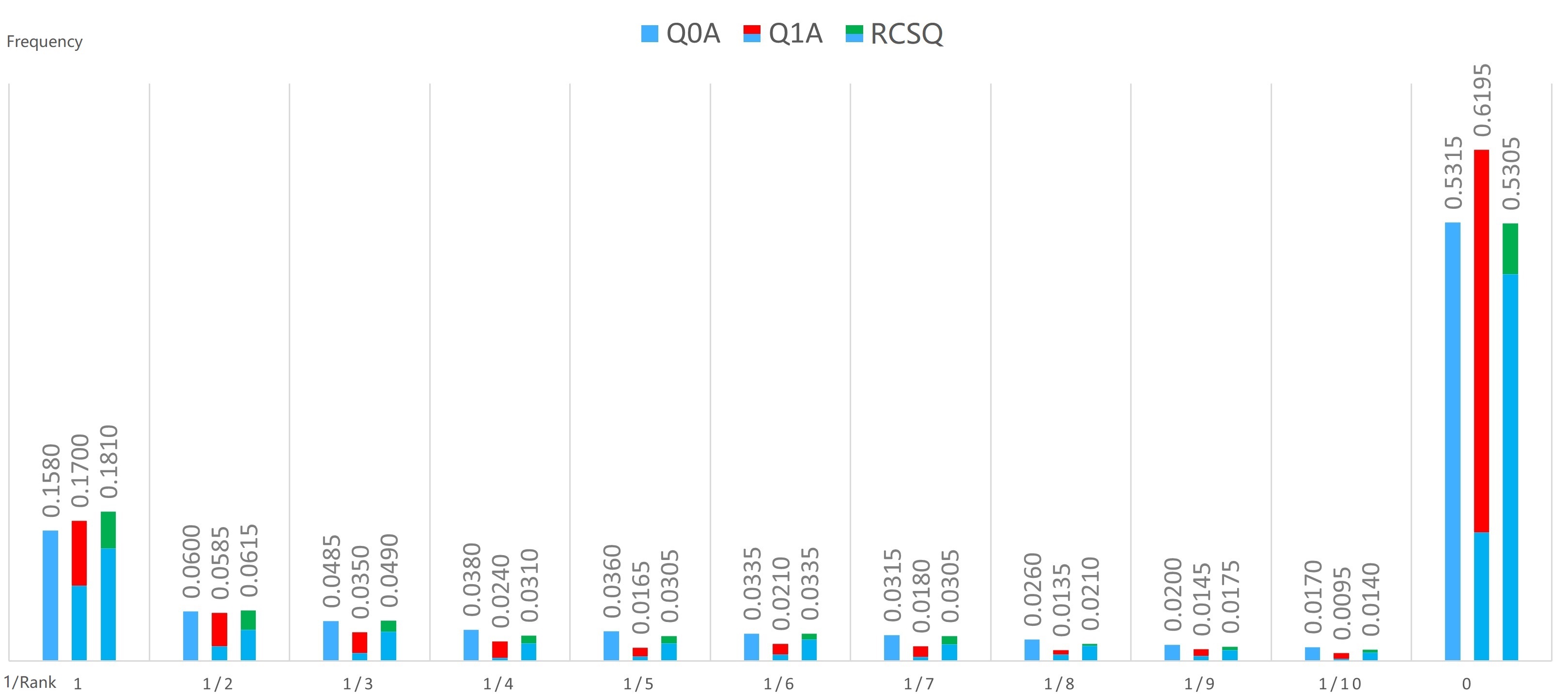}
  \caption{MRR Distributions of our model and baseline models in poly-re-ranker experiment with $\rho=\infty, \tau=2$ on UDC dataset. Blue proportion in Q1A and RCSQ represent the conversations that have the same results as Q0A, red/green proportion represent the conversations that have different result from Q0A, i.e., the improvements.}
  \Description{MRR Distributions for poly-re-ranker experiment and 2 tolerance on UDC dataset.}
 \label{udcdistribution2}
\end{figure}

We also study the MRR distributions as in MSDialog experiments and show the result in Figure~\ref{udcdistribution0} and Figure~\ref{udcdistribution2}. The reciprocal rank distributions in UDC experiments are similar to those in MSDialog experiments except that there are more conversations in group 0. This is because the answer and question re-rankers are less powerful than in MSDialog experiments. We still have the same observations as in MSDialog experiments that the benefit of relevant clarifying questions is offset by irrelevant questions in Q1A baseline, while our model controls the number of irrelevant questions and performs better than both Q0A and Q1A.

Here is an example in UDC dataset of how our model improves the baselines by choosing the better action between answering and asking question.

\begin{itemize}
    \item[\textbf{Case 2}]
    \item[Round 1]
    \item[User:] (Initial Query) \textit{Does anyone know how to force ubuntu to do a hardware-scan? It seems a user's wireless card disappeared.}
    
    \item[Top Ans:] \textit{That sounds like a hardware problem try safe mode.}
    
    \item[Top CQ:] \textit{Does lspci or lsusb (depending on it's hardware) show it ?}

    \item[\textbf{Analysis}:]
    The top retrieved answer is not the correct answer to the query, although it also mentions hardware. The correct answer is ranked in 4th. In this case, Q0A answers with the top answer and thus gets $\text{R@1/100}=0$, MRR=0.25, and Dec.err.=1. 
    
    Q1A, Q2A, and RCSQ will respond to the user with the top retrieved question, which is relevant to the user's query. Hence user will also respond with more information, and the conversation will continue.
    
    \item[Round 2]
    \item[User:] (Reply) \textit{It's in Jigal's court now. We await his response. Thanks for the commands.lspci shows the wireless NIC. Now what should be done?}
    
    \item[Top Ans:] \textit{Check xserver logs in /var/log sorry, don't know.}
    
    \item[Top CQ:] \textit{iwconfig how? it uses the kernel wireless extensions api. read the man pages to figure out how you ask it to, provided your hardware supports that mode. also note, there are caveats with bridging.open a terminal, and run man iwconfig, or in konqueror type man://iwconfig (I think)enjoy =)}
    
     \item[\textbf{Analysis}:]
    The top retrieved answer is again not correct, but thanks to more information, the correct answer is now improved to 2nd. The top retrieved question this time is not relevant to the query. RCSQ avoids risk by choosing to answer with the top retrieved answer, which is the correct decision, thus it gets $\text{R@1/100}=0$, MRR=0.5, and Dec.err.=0. So does Q1A. Q2A asks the top retrieved question, which causes the user to leave the search session. Thus Q2A gets $\text{R@1/100}=0$, MRR=0, and Dec.err.=0.5. In this example, we see that although RCSQ cannot answer the query correctly, it still improves the answer quality and avoids asking irrelevant clarifying question.
\end{itemize}

From both Table~\ref{udcpoly} and Table~\ref{udcbi}, we see that our previous risk control model, which uses both textual and score features, still outperforms the ablation study model variations in almost all the experiments (17 of 18 columns in the two tables). This further strengthens our previous conclusion about using both features. However, in UDC experiments, as user tolerance increases, the text-only model `RCSQ-S' does not become better than the score-only model `RCSQ-T' as it does in MSDialog experiments. We suppose this is because, in UDC conversations, the turn utterance becomes shorter and sparser in meaning, thus the textual features are less useful as in MSDialog experiments.

We now explain another confusing observation in UDC experiments, which differs from MSDialog experiments. In the MSDialog result tables (Table~\ref{msdialogpoly} and Table~\ref{msdialogbi}), the decision error rate of Q0A decreases as the user tolerance increases. We explained the reason earlier that it was because Q0A's decisions are considered less as worse decisions as to the worse decision threshold $1/\tau$ decreases. However, in UDC result tables (Table~\ref{udcpoly} and Table~\ref{udcbi}), the decision error rate of Q0A increases as the user tolerance increases ($0.2870>0.3315>0.3610$, $0.3340>0.3990>0.4315$). The reason can still be found in the definition of `worse decision'. By definition, Q0A's decision to return an irrelevant answer will be worse if the correct answer's reciprocal rank is less than $1/\tau$ and the top retrieved clarifying question is relevant. In MSDialog experiments, the major factor of decision error was the threshold $1/\tau$. However, this is different in UDC experiments. As the user tolerance increases, significantly more relevant clarifying questions are retrieved in UDC experiments, which has a more decisive effect than the decrease of $1/\tau$. This explains why the decision error change of Q0A in UDC experiments differs from that in MSDialog experiments.

\subsection{Opendialkg experiments}

Last, we conduct experiments on the Opendialkg dataset. Opendialkg dataset is originally a conversational reasoning and knowledge graph entity prediction task. The conversation responses are almost impossible to predict without the information in the knowledge graph, because agents are supposed to find entities from the knowledge graph which are relevant to the query and make answers or ask questions around these entities. To make the retrieval task more possible without actually accessing the knowledge graph structure during simulation experiments, we append relevant knowledge graph relationship tuples as texts to each query and conversation turn. After this appending step, we only use the conversations with some knowledge graph information appended and throw the knowledge graph away. During the experiments, we batch conversations and use responses from other conversations in a batch as negative examples to make it a re-ranking task. We hope that the original conversation context, along with the additional knowledge graph information, can be sufficient for the re-ranking of relevant responses.

\begin{table}
\caption{Comparison of all models and baselines on sampled Opendialkg dataset using poly-encoder as re-ranker with 19, 49 and 99 negative samples. $\rho$ is the user patience for total clarifying questions, $\tau$ is the user tolerance for irrelevant clarifying questions. Numbers in bold mean the result is the best excluding oracle. $\dag$ and $\ddag$ means $p < 0.1$ and $p < 0.05$ statistical significance over the best among baseline models.}
\vspace{-0.1in}
\scalebox{0.85}{
\begin{tabular}{l|l|l|l|l|l|l|l|l|l}
\hline\hline

Batchsize & \multicolumn{9}{c}{99 negatives} \\
\hline
\multirow{2}{*}{Users}  & \multicolumn{9}{c}{$\rho=\infty$} \\ 
\cline{2-10}  
& \multicolumn{3}{c|}{$\tau=0$} & \multicolumn{3}{c|}{$\tau=1$} & \multicolumn{3}{c}{$\tau=2$} \\ \hline
Models  & R@1/100     & MRR    & Dec. err    & R@1/100     & MRR    & Dec. err & R@1/100 & MRR & Dec.err \\ \hline
Q0A     & 0.0610 & 0.1127 & 0.4350 &  0.0610 & 0.1127 & 0.5530 & 0.0650 & 0.1179 & 0.6045 \\ 
Q1A   & 0.0510 &  0.0917 & 0.6280   & 0.0710 & 0.1227 & 0.5465 & \textbf{0.0790} & \textbf{0.1359} & 0.5085 \\ 
Q2A     & 0.0170 & 0.0296 & 0.8935  & 0.0285 & 0.0500 & 0.8105 & 0.0355 & 0.0620 & 0.7625  \\ 
CtxPred & 0 & 0.0005 & 0.9966    & 0.0006 & 0.0012 & 0.9943 & 0.0036 &  0.0059 & 0.9691\\ \hline
RCSQ-S     & 0.0675 & 0.1177 & 0.3890 & 0.0720 & \textbf{0.1239}$^{\ddag}$ & 0.4810 & 0.0775 & 0.1325& 0.4940 \\ 
RCSQ-T     & \textbf{0.0700}$^{\ddag}$ & \textbf{0.1207}$^{\ddag}$ & 0.3745 & \textbf{0.0725}$^{\ddag}$ & 0.1235 & 0.5035 & 0.0680 & 0.1190 &  0.5360 \\ 
RCSQ     & 0.0675 & 0.1170 & \textbf{0.3365}$^{\ddag}$ & 0.0685 & 0.1171 & \textbf{0.4635}$^{\ddag}$ & 0.0775 & 0.1335 & \textbf{0.4520}$^{\ddag}$  \\ \hline
Oracle  & 0.0965 & 0.1606 & 0  & 0.1080 & 0.1806 & 0 & 0.1180  & 0.1964 & 0 \\ 
\hline\multirow{2}{*}{Users}  & \multicolumn{9}{c}{$\rho=2$} \\ 
\cline{2-10}  
& \multicolumn{3}{c|}{$\tau=0$} & \multicolumn{3}{c|}{$\tau=1$} & \multicolumn{3}{c}{$\tau=2$} \\ \hline
Models  & R@1/100     & MRR    & Dec. err    & R@1/100     & MRR    & Dec. err & R@1/100 & MRR & Dec.err \\ \hline
Q0A     & \textbf{0.0610} & 0.1127 & 0.4350 &  0.0610 & 0.1127 & 0.5530 & 0.0650 & 0.1179 & 0.6045 \\ 
Q1A   & 0.0510 &  0.0917 & 0.6280   & \textbf{0.0710} & \textbf{0.1227} & 0.5465 & \textbf{0.0790} & \textbf{0.1359} & 0.5085 \\ 
Q2A     & 0.0170 & 0.0296 & 0.8935  & 0.0285 & 0.0500 & 0.8105 & 0.0355 & 0.0620 & 0.7625  \\ 
CtxPred & 0 & 0.0005 & 0.9966    & 0.0006 & 0.0012 & 0.9943 & 0.0036 &  0.0059 & 0.9691\\ \hline
RCSQ-S     & 0.0510 & 0.0924 & 0.6145 &0.0645 & 0.1177 & 0.5340 & 0.0735 & 0.1320& 0.5100 \\ 
RCSQ-T     & 0.0585 & 0.1118 & 0.4085 & 0.0595 & 0.1140 & 0.4895 & 0.0620 & 0.1157 &  0.5675 \\ 
RCSQ     & 0.0585 & \textbf{0.1129} & \textbf{0.3875}$^{\ddag}$ &  0.0665& 0.1212 & \textbf{0.4380}$^{\ddag}$ & 0.0746 & 0.1310 & \textbf{0.4682}$^{\ddag}$  \\ \hline
Oracle  & 0.0995 & 0.1653 & 0  & 0.1190 & 0.1924 & 0 & 0.1281  & 0.2055 & 0 \\ 
\hline\hline
Batchsize & \multicolumn{9}{c}{49 negatives} \\
\hline
\multirow{2}{*}{Users}  & \multicolumn{9}{c}{$\rho=\infty$} \\ 
\cline{2-10}  
   &  \multicolumn{3}{c|}{$\tau=0$} & \multicolumn{3}{c|}{$\tau=1$} & \multicolumn{3}{c}{$\tau=2$}\\ \hline
Models  & R@1/50      & MRR    & Dec. err    & R@1/50     & MRR    & Dec. err & R@1/50  & MRR & Dec.err \\ \hline
Q0A & 0.0920 & 0.1718 & 0.5040    & 0.0920 & 0.1718 & 0.6040 & 0.0920 & 0.1718 & 0.6385  \\ 
Q1A  & 0.0965  & 0.1579  & 0.5720 & 0.1160 & 0.1936 & 0.4995 & 0.1240 & 0.2097 & 0.4605 \\ 
Q2A     & 0.0365  &  0.0581   & 0.8460    & 0.0600 & 0.0959  & 0.7405 & 0.0750 & 0.1209 & 0.6675  \\ 
CtxPred &  0.0015   & 0.0021   & 0.9925  & 0.0045 & 0.0066 & 0.9830 & 0.0055 & 0.0077 & 0.9840  \\ \hline
RCSQ-S  &  \textbf{0.1015}$^{\ddag}$ & 0.1679 & 0.5225 & 0.1155 & 0.1935 & 0.5010 & \textbf{0.1230}$^{\ddag}$ & 0.2088 & 0.4610  \\ 
RCSQ-T  & 0.0950&  \textbf{0.1736}$^{\ddag}$ & 0.4920 & 0.1025 & 0.1811 & 0.5495 & 0.1050 & 0.1856 & 0.5700  \\
RCSQ     & 0.1010  & 0.1730 &  \textbf{0.4395}$^{\ddag}$ & \textbf{0.1170}$^{\ddag}$ & \textbf{0.1989}$^{\ddag}$ & \textbf{0.4425}$^{\ddag}$ & 0.1225 & \textbf{0.2097}$^{\ddag}$ & \textbf{0.4385}$^{\ddag}$  \\ \hline
Oracle  & 0.1575  & 0.2533  & 0   & 0.1810 & 0.2868 & 0 & 0.1955 & 0.3054 & 0.0300\\ 
\hline\hline
Batchsize & \multicolumn{9}{c}{19 negatives} \\
\hline
\multirow{2}{*}{Users}  & \multicolumn{9}{c}{$\rho=\infty$} \\ 
\cline{2-10}  

&  \multicolumn{3}{c|}{$\tau=0$} & \multicolumn{3}{c|}{$\tau=1$} & \multicolumn{3}{c}{$\tau=2$}\\ \hline
Models  & R@1/20     & MRR    & Dec. err    & R@1/20      & MRR    & Dec. err & R@1/20  & MRR & Dec.err \\ \hline
Q0A     &  0.1590    & 0.2964    & 0.6120  & 0.1590  & 0.2964  & 0.6375 & 0.1590    & 0.2964  & 0.6040 \\ 
Q1A   & 0.1875    &  0.3009  & 0.5180    & 0.2210 & 0.3519 & 0.4445& 0.2285 & 0.3677 & 0.4095 \\ 
Q2A     &  0.1075   & 0.1605    & 0.6900    &  0.1515&  0.2258  & 0.5640 & 0.1720 & 0.2633 & 0.4775   \\
CtxPred &  0.0435   & 0.0675   & 0.8785 & 0.0620& 0.0938 &  0.8405 & 0.0755 & 0.1194 & 0.7655 \\ \hline
RCSQ-S     &0.1885 & 0.3045 & 0.5055 & 0.2205 & 0.3509 & 0.4440 & \textbf{0.2270}$^{\ddag}$ & \textbf{0.3658}$^{\ddag}$ & \textbf{0.4100}$^{\ddag}$ \\ 
RCSQ-T     & 0.1790 & \textbf{0.3114}$^{\ddag}$ & 0.5285 & 0.1910 &0.3255  & 0.5035 &  0.1935 & 0.3288 &  0.5065 \\ 
RCSQ     & \textbf{0.1905}$^{\ddag}$ & 0.3106 & \textbf{0.4620}$^{\ddag}$ & \textbf{0.2240}$^{\ddag}$& \textbf{0.3547}$^{\ddag}$ &  \textbf{0.4235}$^{\ddag}$& 0.2210 & 0.3620 & 0.4140  \\ \hline
Oracle  & 0.2815    & 0.4304       & 0    & 0.3165  & 0.4669 & 0.0460 & 0.3310 & 0.4858 & 0  \\ 
\hline\hline
\end{tabular}
}
\label{openpoly}
\end{table}

\begin{table}
\caption{Comparison of all models and baselines on sampled Opendialkg dataset using bi-encoder as re-ranker with 19, 49 and 99 negative samples. $\rho$ is the user patience for total clarifying questions, $\tau$ is the user tolerance for irrelevant clarifying questions. Numbers in bold mean the result is the best excluding oracle. $\dag$ and $\ddag$ means $p < 0.1$ and $p < 0.05$ statistical significance over the best among baseline models.}
\vspace{-0.1in}
\scalebox{0.85}{
\begin{tabular}{l|l|l|l|l|l|l|l|l|l}
\hline\hline
Batchsize & \multicolumn{9}{c}{99 negatives} \\
\hline
\multirow{2}{*}{Users}  & \multicolumn{9}{c}{$\rho=\infty$} \\ 
\cline{2-10} & \multicolumn{3}{c|}{$\tau=0$} & \multicolumn{3}{c|}{$\tau=1$} & \multicolumn{3}{c}{$\tau=2$} \\ \hline
Models  & R@1/100    & MRR    & Dec. err    & R@1/100     & MRR    & Dec. err & R@1/100 & MRR & Dec.err \\ \hline
Q0A     & 0.0535 &0.1022 & 0.4350 &  0.0535 & 0.1022 & 0.4480 & 0.0535 & 0.1022 & 0.5100 \\ 
Q1A   & 0.0430 &  0.0751 & 0.6950  & 0.0515 & 0.0940 & 0.6300 & 0.0620 & 0.1087 & 0.5760 \\ 
Q2A     & 0.0160 & 0.0232 & 0.9175  & 0.0250 & 0.0388 & 0.8530 & 0.0290 & 0.0484 & 0.8030  \\ 
CtxPred & 0 & 0.0005 & 0.9966    & 0.0006 & 0.0012 & 0.9943 & 0.0036 &  0.0059 & 0.9691\\ \hline
RCSQ-S  & 0.0555 &0.1057 &0.3180&  0.0590& 0.1003& 0.4760& \textbf{0.0610}$^{\ddag}$& 0.1068 &0.4895\\ 
RCSQ-T  & 0.0565& 0.1050& 0.3060&  0.0580& \textbf{0.1075}$^{\ddag}$ &0.4045& 0.0590& \textbf{0.1078}$^{\ddag}$ &0.4555 \\ 
RCSQ & \textbf{0.0580}$^{\ddag}$&\textbf{0.1079}$^{\ddag}$& \textbf{0.2745}$^{\ddag}$& \textbf{0.0595}$^{\ddag}$& 0.1066& \textbf{0.3875}$^{\ddag}$&  0.0595& 0.1071 &\textbf{0.4405}$^{\ddag}$\\ \hline
Oracle  & 0.0785 &0.1363& 0& 0.0895& 0.1525&0& 0.0995& 0.1666& 0\\ 
\hline\multirow{2}{*}{Users}  & \multicolumn{9}{c}{$\rho=2$} \\ 
\cline{2-10}  
& \multicolumn{3}{c|}{$\tau=0$} & \multicolumn{3}{c|}{$\tau=1$} & \multicolumn{3}{c}{$\tau=2$} \\ \hline
Models  & R@1/100     & MRR    & Dec. err    & R@1/100     & MRR    & Dec. err & R@1/100 & MRR & Dec.err \\ \hline
Q0A     & 0.0535 &0.1022 & 0.4350 &  0.0535 & 0.1022 & 0.4480 & 0.0535 & 0.1022 & 0.5100 \\ 
Q1A   & 0.0430 &  0.0751 & 0.6950  & 0.0515 & 0.0940 & 0.6300 & 0.0620 & 0.1087 & 0.5760 \\ 
Q2A     & 0.0160 & 0.0232 & 0.9175  & 0.0250 & 0.0388 & 0.8530 & 0.0290 & 0.0484 & 0.8030  \\ 
CtxPred & 0 & 0.0005 & 0.9966    & 0.0006 & 0.0012 & 0.9943 & 0.0036 &  0.0059 & 0.9691\\ \hline
RCSQ-S     & 0.0535 & 0.1022 & 0.3490 &0.0580 & 0.1037 & 0.4272 & 0.0625 &  \textbf{0.1086}$^{\ddag}$& 0.5660 \\ 
RCSQ-T     & \textbf{0.0555}$^{\ddag}$ & \textbf{0.1037}$^{\ddag}$ &  0.4272 & \textbf{0.0595}$^{\ddag}$ & \textbf{0.1140}$^{\ddag}$ & 0.4895 & 0.0540 & 0.1030 &  0.5060 \\ 
RCSQ     & 0.0545 & 0.0992 & \textbf{0.3410}$^{\ddag}$ &  0.0580& 0.1061 & \textbf{0.4090}$^{\ddag}$ &  \textbf{0.0641}$^{\ddag}$ & 0.1081 & \textbf{0.5351}$^{\ddag}$  \\ \hline
Oracle  & 0.0780 & 0.1390 & 0  & 0.0885 & 0.1540 & 0 & 0.0977  & 0.1654 & 0 \\ 
\hline\hline
Batchsize & \multicolumn{9}{c}{49 negatives} \\
\hline
\multirow{2}{*}{Users}  & \multicolumn{9}{c}{$\rho=\infty$} \\ 
\cline{2-10}  

   &  \multicolumn{3}{c|}{$\tau=0$} & \multicolumn{3}{c|}{$\tau=1$} & \multicolumn{3}{c}{$\tau=2$}\\ \hline
Models  & R@1/50    & MRR    & Dec. err    & R@1/50     & MRR    & Dec. err & R@1/50 & MRR & Dec.err \\ \hline
Q0A & 0.0840 & 0.1635 &0.4130& 0.0840& 0.1635& 0.5110&  0.0840 &0.1635 &0.5655  \\ 
Q1A & 0.0735& 0.1251& 0.6400 & 0.0920& 0.1616 &0.5445& 0.1060& 0.1834 &0.5035 \\ 
Q2A & 0.0325 &0.0489 &0.8775& 0.0580& 0.0845 &0.7920& 0.0710 &0.1084 &0.7145 \\ 
CtxPred & 0.0050& 0.0079& 0.9815& 0& 0.0002& 1& 0.0015& 0.0035 &0.9895\\ \hline
RCSQ-S  & 0.0875 &0.1585 &0.4045 & 0.0925& 0.1621& 0.5475& \textbf{0.1065}$^{\ddag}$ &0.1831 &0.5075\\ 
RCSQ-T  & 0.0845 &\textbf{0.1641}$^{\ddag}$ &0.3980 & 0.0890& \textbf{0.1676}$^{\ddag}$ &0.4760& 0.0900 &0.1701 &0.5375 \\
RCSQ &\textbf{0.0900}$^{\ddag}$ &0.1640& \textbf{0.3535}$^{\ddag}$ & \textbf{0.0925}$^{\ddag}$& 0.1672& \textbf{0.4425}$^{\ddag}$ & 0.1060 &\textbf{0.1849}$^{\ddag}$& \textbf{0.4705}$^{\ddag}$ \\ \hline
Oracle  & 0.1310 &0.2239 &0& 0.1570 &0.2540 &0& 0.1770& 0.2771 &0\\ 
\hline\hline

Batchsize & \multicolumn{9}{c}{19 negatives} \\
\hline
\multirow{2}{*}{Users}  & \multicolumn{9}{c}{$\rho=\infty$} \\ 
\cline{2-10}  
  &  \multicolumn{3}{c|}{$\tau=0$} & \multicolumn{3}{c|}{$\tau=1$} & \multicolumn{3}{c}{$\tau=2$}\\ \hline
Models  & R@1/20     & MRR    & Dec. err    & R@1/20     & MRR    & Dec. err & R@1/20 & MRR & Dec.err \\ \hline
Q0A     & 0.1630 & 0.3004 & 0.4965& 0.1630 & 0.3004 & 0.5670 & 0.1630 & 0.3004 & 0.5565\\ 
Q1A     & \textbf{0.1810} & 0.2719 & 0.5480& 0.2225 & 0.3401 & 0.4525& \textbf{0.2375} & \textbf{0.3678} & 0.4040 \\ 
Q2A     & 0.0880 & 0.1257 & 0.7780& 0.1330 & 0.1951 & 0.6405 & 0.1585 &  0.2387 & 0.5430\\
CtxPred & 0 & 0 & 1 & 0 & 0 & 1 & 0.0665 & 0.0999 & 0.8210\\ \hline
RCSQ-S  & 0.1655 & 0.3023 & 0.4835& 0.2220 & 0.3389 & 0.4550&  0.2370 & 0.3673 & 0.4035\\ 
RCSQ-T  & 0.1670 & 0.3022 & 0.4860& 0.1680&  0.3055 & 0.5510 & 0.1710 & 0.3087 & 0.5400\\ 
RCSQ & 0.1790 & \textbf{0.3118}$^{\ddag}$ & \textbf{0.4310}$^{\ddag}$ & \textbf{0.2245}$^{\ddag}$ & \textbf{0.3457}$^{\ddag}$ & \textbf{0.4345}$^{\ddag}$ & 0.2370 & 0.3674 & \textbf{0.3995}$^{\ddag}$ \\ \hline
Oracle  & 0.2620 & 0.4055 & 0& 0.3015 & 0.4491 & 0 & 0.3265 & 0.4781 & 0 \\ 
\hline\hline
\end{tabular}
}
\label{openbi}
\end{table}

In Table~\ref{openpoly} and Table~\ref{openbi}, we show the results of our risk control model combined with poly-encoder and bi-encoder, respectively. When we compare the performances of the baselines in Opendialkg experiments with 99 negatives with their performances in MSDialog/UDC experiments, the baselines are significantly worse in Opendialkg. The main reason is that appending parts of knowledge graph as texts to conversation turns is not an ideal solution to knowledge graph involved conversation reasoning. Other reasons include that the conversations in the Opendialkg dataset are closer to chats rather than question answering threads, which means they do not always have clear information needs from the beginning. Without clear information needs from the beginning, Q0A will have no clue of how the conversation ends (which is the true answer to the conversation), and Q1A will have no clue of what to ask. Because of these reasons and the low performances, we make the task slightly easier by reducing the batch size from 100 to 50 and 20, which means the answer/question re-rankers will have fewer candidates to re-rank. The performances are still not ideal, so we stop changing the batch size and conclude that it is very hard to use this dataset for our simulation experiments. However, the results we get from these experiments can still strengthen some of our previous conclusions that our model can improve the baselines.

Here is an example in the Opendialkg dataset of how our model improves the baselines by choosing the better action between answering and asking questions. In opendialkg, the conversations usually have multiple questions and hence the retrieved clarifying questions are not always questions but intermediate answers to the user's query. From this example, we can also see why our model is not ideal for opendialkg, since the dataset has too many session search conversations where users ask the system multiple questions one after another. This is not the scenario our system is supposed to solve, as explained in the Related Work section.

\begin{itemize}
    \item[\textbf{Case 3}]
    \item[Round 1]
    \item[User:] (Initial Query) \textit{I really liked the movie As Good As It Gets. Who starred in that?}
    
    \item[Top Ans:] \textit{I Do, Jennifer Connelly, Russell Crowe and Ed Harris starred in the movie A Beautiful Mind.}
    
    \item[Top CQ:] (Intermediate answer) \textit{Great movie, some of the actors include Yeardley Smity, Wood Harris, and Julie Benz.}

    \item[\textbf{Analysis}:]
    The top retrieved answer is terrible since it does not even answer the query. Actually, the correct answer is ranked in 6th. In this case, Q0A answers with the top answer and thus gets $\text{R@1/100}=0$, MRR=0.16, and Dec.err.=1. 
    
    Q1A, Q2A, and RCSQ respond the user with the top result from the question retriever, which is an intermediate answer to the user's query. Hence user will continue to interact with the system.
    
    \item[Round 2]
    \item[User:] (Ask again) \textit{That name sounds familiar. What else has Yeardley Smith starred in?}
    
    \item[Top Ans:] \textit{I Do, Jennifer Connelly, Russell Crowe and Ed Harris starred in the movie A Beautiful Mind.}
    
    \item[Top CQ:] (Intermediate answer)  \textit{Yeardley Smith also starred in Virginia which is a romantic film.}
    
    \item[\textbf{Analysis}:]
    The top retrieved answer is again not correct. Q1A will answer the query with the same wrong answer, but now the correct answer ranking is improved to 4th thanks to more context, thus Q1A gets $\text{R@1/100}=0$, MRR=0.25, and Dec.err.=0.5 (Correct in round 1, then wrong in round 2). 
    
    RCSQ and Q2A choose to respond with the intermediate answer retrieved by question retriever, and user will respond again and continue the conversation.
    
    \item[Round 3]
    \item[User:] (Ask again) \textit{I haven't seen that one. Was she also in City Slickers?}
    
    \item[Top Ans:] \textit{Meat Loaf stars in Fight Club.  He also stars in Wayne's World and The 51st State.}
    
    \item[Top CQ:] (Intermediate answer)  \textit{Reg E. Cathey also starred in The Machinist, and Tank Girl.}
    
    \item[\textbf{Analysis}:]
    The top retrieved answer is still not correct, but the correct answer ranking further improves to 3rd. Here RSCQ chooses to answer with top retrieved answer and gets $\text{R@1/100}=0$, MRR=0.33, and Dec.err.=1 (Correct in all 3 rounds). This is another example besides the UDC example, which shows that although RCSQ cannot guarantee to retrieve correct answer, it still makes the best decisions to improve answer quality and avoid asking irrelevant questions.  
    
\end{itemize}

We compare the performances of our models and the baseline models in these experiments and find that our risk control model and its ablation study variations still outperform the baseline models in most experiments (51 of 54 columns in Table~\ref{openpoly} and Table~\ref{openbi}). In these experiments, our model outperforms ablation study variations in most experiments (36 of 54 columns in the two tables). This further strengthens our previous conclusions that using the risk control model with the retrieval models is a better strategy, and having both textual features and score features as input of the risk control model is the best.

Although our experiments on Opendialkg dataset show that our model, including ablation study variations, is better than the baseline models in most experiments. We consider our model not an ideal solution to Opendialkg dataset, and our results should not be compared seriously beyond our experiments. There are multiple reasons for that. The most important reason is that we do not fully make use of the knowledge graph of the Opendialkg dataset in our experiments. Ideally, the best solution to knowledge graph-involved tasks is to make use of the knowledge graph in the model by building individual models for the graph. It will be interesting to have both answer and clarifying question generators that use the conversation context and the knowledge graph to solve this task, and we are looking forward to testing our risk control model with such generators in the future.

Another reason is that our assumptions about clarifying question order do not hold on the Opendialkg dataset. In Opendialkg conversations, the user often starts with a simple question and develops additional information needs after system responses. This invalidates both of our assumptions that 1) a clarifying question in later turns is sill valid question and 2) clarifying question order can be changed during the conversation. These make the clarifying re-ranker more unreliable compared to MSDialog and UDC experiments. As a result, the clarifying question scores output by the clarifying question re-ranker become more unreliable features, which harms the performance of our framework.

\section{Conclusions}

In this paper, we proposed the concept of risk in asking clarifying questions to users in conversational search. We find that although existing methods study the benefit of asking clarifying questions in conversational search and methodologies of generating them; they overlook the risk in asking clarifying questions that can decrease answer quality and harm user experience. We propose that asking clarifying questions should not be always taken as the default alternative to answering the query since irrelevant questions are equally bad and often worse than suboptimal answers. To control such risks, we revise our previously proposed risk-aware conversational search framework and a risk control model which comprehensively evaluates conversation contexts and the best retrievable answers and clarifying questions before deciding whether to answer the query or ask a clarifying question. Through extensive simulation experiments on three conversation datasets and with different retrieval models, we show that our risk control model can improve answer quality and user's search experience. Through detailed analyses and case studies, we verify the improvements of our model over the baseline models are meaningful and provide new insights and explanations of why our proposed model outperforms the baseline models.

We consider multiple future work directions of our work. We only experiment with our model together with response retrieval models in this work. However, it will be interesting to further test our model with response generation models or retrieval involving knowledge graphs. In this work, we make some assumptions about clarifying questions being valid at different conversation turns and order-independent. It will be interesting to assume otherwise and essentially further increase the risk of asking clarifying questions. In this work, we conduct experiments with different user models but we only assume a singleton user model at a time, which is not realistic in real-world search scenarios. Therefore, another promising direction is to develop simulation and evaluation settings to support different user types.

\begin{acks}
This work was supported in part by the School of Computing, University of Utah. Any opinions, findings and conclusions, or recommendations expressed in this material are those of the authors and do not necessarily reflect those of the sponsor.
\end{acks}

\bibliographystyle{ACM-Reference-Format}
\bibliography{sample-base}


\begin{thebibliography}{45}


\ifx \showCODEN    \undefined \def \showCODEN     #1{\unskip}     \fi
\ifx \showDOI      \undefined \def \showDOI       #1{#1}\fi
\ifx \showISBNx    \undefined \def \showISBNx     #1{\unskip}     \fi
\ifx \showISBNxiii \undefined \def \showISBNxiii  #1{\unskip}     \fi
\ifx \showISSN     \undefined \def \showISSN      #1{\unskip}     \fi
\ifx \showLCCN     \undefined \def \showLCCN      #1{\unskip}     \fi
\ifx \shownote     \undefined \def \shownote      #1{#1}          \fi
\ifx \showarticletitle \undefined \def \showarticletitle #1{#1}   \fi
\ifx \showURL      \undefined \def \showURL       {\relax}        \fi
\providecommand\bibfield[2]{#2}
\providecommand\bibinfo[2]{#2}
\providecommand\natexlab[1]{#1}
\providecommand\showeprint[2][]{arXiv:#2}

\bibitem[\protect\citeauthoryear{Aliannejadi, Chakraborty, R{\'\i}ssola, and
  Crestani}{Aliannejadi et~al\mbox{.}}{2020a}]%
        {relevantturn}
\bibfield{author}{\bibinfo{person}{Mohammad Aliannejadi},
  \bibinfo{person}{Manajit Chakraborty}, \bibinfo{person}{Esteban~Andr{\'e}s
  R{\'\i}ssola}, {and} \bibinfo{person}{Fabio Crestani}.}
  \bibinfo{year}{2020}\natexlab{a}.
\newblock \showarticletitle{Harnessing evolution of multi-turn conversations
  for effective answer retrieval}. In \bibinfo{booktitle}{\emph{Proceedings of
  the 2020 Conference on Human Information Interaction and Retrieval}}.
  \bibinfo{pages}{33--42}.
\newblock


\bibitem[\protect\citeauthoryear{Aliannejadi, Kiseleva, Chuklin, Dalton, and
  Burtsev}{Aliannejadi et~al\mbox{.}}{2020b}]%
        {aliannejadi2020convai3}
\bibfield{author}{\bibinfo{person}{Mohammad Aliannejadi},
  \bibinfo{person}{Julia Kiseleva}, \bibinfo{person}{Aleksandr Chuklin},
  \bibinfo{person}{Jeff Dalton}, {and} \bibinfo{person}{Mikhail Burtsev}.}
  \bibinfo{year}{2020}\natexlab{b}.
\newblock \showarticletitle{ConvAI3: Generating Clarifying Questions for
  Open-Domain Dialogue Systems (ClariQ)}.
\newblock \bibinfo{journal}{\emph{arXiv preprint arXiv:2009.11352}}
  (\bibinfo{year}{2020}).
\newblock


\bibitem[\protect\citeauthoryear{Aliannejadi, Zamani, Crestani, and
  Croft}{Aliannejadi et~al\mbox{.}}{2019}]%
        {askingcq}
\bibfield{author}{\bibinfo{person}{Mohammad Aliannejadi},
  \bibinfo{person}{Hamed Zamani}, \bibinfo{person}{Fabio Crestani}, {and}
  \bibinfo{person}{W.~Bruce Croft}.} \bibinfo{year}{2019}\natexlab{}.
\newblock \showarticletitle{Asking Clarifying Questions in Open-Domain
  Information-Seeking Conversations}.
\newblock \bibinfo{journal}{\emph{CoRR}}  \bibinfo{volume}{abs/1907.06554}
  (\bibinfo{year}{2019}).
\newblock
\showeprint[arxiv]{1907.06554}
\urldef\tempurl%
\url{http://arxiv.org/abs/1907.06554}
\showURL{%
\tempurl}


\bibitem[\protect\citeauthoryear{Allan}{Allan}{2005}]%
        {trechard}
\bibfield{author}{\bibinfo{person}{James Allan}.}
  \bibinfo{year}{2005}\natexlab{}.
\newblock \bibinfo{booktitle}{\emph{HARD track overview in TREC 2003 high
  accuracy retrieval from documents}}.
\newblock \bibinfo{type}{{T}echnical {R}eport}.
  \bibinfo{institution}{MASSACHUSETTS UNIV AMHERST CENTER FOR INTELLIGENT
  INFORMATION RETRIEVAL}.
\newblock


\bibitem[\protect\citeauthoryear{Bar-Yossef and Kraus}{Bar-Yossef and
  Kraus}{2011}]%
        {querycompletionwithcontext}
\bibfield{author}{\bibinfo{person}{Ziv Bar-Yossef} {and} \bibinfo{person}{Naama
  Kraus}.} \bibinfo{year}{2011}\natexlab{}.
\newblock \showarticletitle{Context-sensitive query auto-completion}. In
  \bibinfo{booktitle}{\emph{Proceedings of the 20th international conference on
  World wide web}}. \bibinfo{pages}{107--116}.
\newblock


\bibitem[\protect\citeauthoryear{Bi, Ai, Zhang, and Croft}{Bi
  et~al\mbox{.}}{2019}]%
        {negativefeedback}
\bibfield{author}{\bibinfo{person}{Keping Bi}, \bibinfo{person}{Qingyao Ai},
  \bibinfo{person}{Yongfeng Zhang}, {and} \bibinfo{person}{W~Bruce Croft}.}
  \bibinfo{year}{2019}\natexlab{}.
\newblock \showarticletitle{Conversational product search based on negative
  feedback}. In \bibinfo{booktitle}{\emph{Proceedings of the 28th ACM
  International Conference on Information and Knowledge Management}}.
  \bibinfo{pages}{359--368}.
\newblock


\bibitem[\protect\citeauthoryear{Boldi, Bonchi, Castillo, Donato, Gionis, and
  Vigna}{Boldi et~al\mbox{.}}{2008}]%
        {boldi2008}
\bibfield{author}{\bibinfo{person}{Paolo Boldi}, \bibinfo{person}{Francesco
  Bonchi}, \bibinfo{person}{Carlos Castillo}, \bibinfo{person}{Debora Donato},
  \bibinfo{person}{Aristides Gionis}, {and} \bibinfo{person}{Sebastiano
  Vigna}.} \bibinfo{year}{2008}\natexlab{}.
\newblock \showarticletitle{The query-flow graph: model and applications}. In
  \bibinfo{booktitle}{\emph{Proceedings of the 17th ACM conference on
  Information and knowledge management}}. \bibinfo{pages}{609--618}.
\newblock


\bibitem[\protect\citeauthoryear{Boldi, Bonchi, Castillo, Donato, and
  Vigna}{Boldi et~al\mbox{.}}{2009}]%
        {boldi2009}
\bibfield{author}{\bibinfo{person}{Paolo Boldi}, \bibinfo{person}{Francesco
  Bonchi}, \bibinfo{person}{Carlos Castillo}, \bibinfo{person}{Debora Donato},
  {and} \bibinfo{person}{Sebastiano Vigna}.} \bibinfo{year}{2009}\natexlab{}.
\newblock \showarticletitle{Query suggestions using query-flow graphs}. In
  \bibinfo{booktitle}{\emph{Proceedings of the 2009 workshop on Web Search
  Click Data}}. \bibinfo{pages}{56--63}.
\newblock


\bibitem[\protect\citeauthoryear{Braslavski, Savenkov, Agichtein, and
  Dubatovka}{Braslavski et~al\mbox{.}}{2017}]%
        {whatdoyoumean}
\bibfield{author}{\bibinfo{person}{Pavel Braslavski}, \bibinfo{person}{Denis
  Savenkov}, \bibinfo{person}{Eugene Agichtein}, {and} \bibinfo{person}{Alina
  Dubatovka}.} \bibinfo{year}{2017}\natexlab{}.
\newblock \showarticletitle{What do you mean exactly? Analyzing clarification
  questions in CQA}. In \bibinfo{booktitle}{\emph{Proceedings of the 2017
  Conference on Conference Human Information Interaction and Retrieval}}.
  \bibinfo{pages}{345--348}.
\newblock


\bibitem[\protect\citeauthoryear{Cao, Jiang, Pei, He, Liao, Chen, and Li}{Cao
  et~al\mbox{.}}{2008}]%
        {querysuggestion2}
\bibfield{author}{\bibinfo{person}{Huanhuan Cao}, \bibinfo{person}{Daxin
  Jiang}, \bibinfo{person}{Jian Pei}, \bibinfo{person}{Qi He},
  \bibinfo{person}{Zhen Liao}, \bibinfo{person}{Enhong Chen}, {and}
  \bibinfo{person}{Hang Li}.} \bibinfo{year}{2008}\natexlab{}.
\newblock \showarticletitle{Context-aware query suggestion by mining
  click-through and session data}. In \bibinfo{booktitle}{\emph{Proceedings of
  the 14th ACM SIGKDD international conference on Knowledge discovery and data
  mining}}. \bibinfo{pages}{875--883}.
\newblock


\bibitem[\protect\citeauthoryear{Chen and Wang}{Chen and Wang}{2019}]%
        {chen2019sequentialonUDC}
\bibfield{author}{\bibinfo{person}{Qian Chen} {and} \bibinfo{person}{Wen
  Wang}.} \bibinfo{year}{2019}\natexlab{}.
\newblock \showarticletitle{Sequential attention-based network for noetic
  end-to-end response selection}.
\newblock \bibinfo{journal}{\emph{arXiv preprint arXiv:1901.02609}}
  (\bibinfo{year}{2019}).
\newblock


\bibitem[\protect\citeauthoryear{Cho, Zhang, Rao, Brockett, and Lee}{Cho
  et~al\mbox{.}}{2019}]%
        {askingcqnlgmulti}
\bibfield{author}{\bibinfo{person}{Woon~Sang Cho}, \bibinfo{person}{Yizhe
  Zhang}, \bibinfo{person}{Sudha Rao}, \bibinfo{person}{Chris Brockett}, {and}
  \bibinfo{person}{Sungjin Lee}.} \bibinfo{year}{2019}\natexlab{}.
\newblock \showarticletitle{Generating a Common Question from Multiple
  Documents using Multi-source Encoder-Decoder Models}.
\newblock \bibinfo{journal}{\emph{arXiv preprint arXiv:1910.11483}}
  (\bibinfo{year}{2019}).
\newblock


\bibitem[\protect\citeauthoryear{Coden, Gruhl, Lewis, and Mendes}{Coden
  et~al\mbox{.}}{2015}]%
        {aorb}
\bibfield{author}{\bibinfo{person}{Anni Coden}, \bibinfo{person}{Daniel Gruhl},
  \bibinfo{person}{Neal Lewis}, {and} \bibinfo{person}{Pablo~N Mendes}.}
  \bibinfo{year}{2015}\natexlab{}.
\newblock \showarticletitle{Did you mean A or B? Supporting Clarification
  Dialog for Entity Disambiguation.}. In \bibinfo{booktitle}{\emph{SumPre-HSWI@
  ESWC}}.
\newblock


\bibitem[\protect\citeauthoryear{Devlin, Chang, Lee, and Toutanova}{Devlin
  et~al\mbox{.}}{2018}]%
        {bert}
\bibfield{author}{\bibinfo{person}{Jacob Devlin}, \bibinfo{person}{Ming-Wei
  Chang}, \bibinfo{person}{Kenton Lee}, {and} \bibinfo{person}{Kristina
  Toutanova}.} \bibinfo{year}{2018}\natexlab{}.
\newblock \showarticletitle{Bert: Pre-training of deep bidirectional
  transformers for language understanding}.
\newblock \bibinfo{journal}{\emph{arXiv preprint arXiv:1810.04805}}
  (\bibinfo{year}{2018}).
\newblock


\bibitem[\protect\citeauthoryear{Emanuel}{Emanuel}{1968}]%
        {emanuel1968sequencing}
\bibfield{author}{\bibinfo{person}{A~Schegloff Emanuel}.}
  \bibinfo{year}{1968}\natexlab{}.
\newblock \showarticletitle{Sequencing in conversational openings}.
\newblock \bibinfo{journal}{\emph{American Anthropologist}}
  \bibinfo{volume}{70}, \bibinfo{number}{6} (\bibinfo{year}{1968}),
  \bibinfo{pages}{1075--1095}.
\newblock


\bibitem[\protect\citeauthoryear{Henderson, Casanueva, Mrk{\v{s}}i{\'c}, Su,
  Wen, and Vuli{\'c}}{Henderson et~al\mbox{.}}{2019}]%
        {henderson2019convertonUDC}
\bibfield{author}{\bibinfo{person}{Matthew Henderson}, \bibinfo{person}{Inigo
  Casanueva}, \bibinfo{person}{Nikola Mrk{\v{s}}i{\'c}},
  \bibinfo{person}{Pei-Hao Su}, \bibinfo{person}{Tsung-Hsien Wen}, {and}
  \bibinfo{person}{Ivan Vuli{\'c}}.} \bibinfo{year}{2019}\natexlab{}.
\newblock \showarticletitle{Convert: Efficient and accurate conversational
  representations from transformers}.
\newblock \bibinfo{journal}{\emph{arXiv preprint arXiv:1911.03688}}
  (\bibinfo{year}{2019}).
\newblock


\bibitem[\protect\citeauthoryear{Hu, Zhang, Bolivar, and Shoup}{Hu
  et~al\mbox{.}}{2019}]%
        {autocomplete}
\bibfield{author}{\bibinfo{person}{Wenyan Hu}, \bibinfo{person}{Xiaodi Zhang},
  \bibinfo{person}{Alvaro Bolivar}, {and} \bibinfo{person}{Randall~Scott
  Shoup}.} \bibinfo{year}{2019}\natexlab{}.
\newblock \bibinfo{title}{Search box auto-complete}.
\newblock
\newblock
\newblock
\shownote{US Patent App. 16/237,245.}


\bibitem[\protect\citeauthoryear{Humeau, Shuster, Lachaux, and Weston}{Humeau
  et~al\mbox{.}}{2019}]%
        {polyencoder}
\bibfield{author}{\bibinfo{person}{Samuel Humeau}, \bibinfo{person}{Kurt
  Shuster}, \bibinfo{person}{Marie-Anne Lachaux}, {and} \bibinfo{person}{Jason
  Weston}.} \bibinfo{year}{2019}\natexlab{}.
\newblock \showarticletitle{Poly-encoders: Transformer architectures and
  pre-training strategies for fast and accurate multi-sentence scoring}.
\newblock \bibinfo{journal}{\emph{arXiv preprint arXiv:1905.01969}}
  (\bibinfo{year}{2019}).
\newblock


\bibitem[\protect\citeauthoryear{Kaiser, Roy, and Weikum}{Kaiser
  et~al\mbox{.}}{2020}]%
        {wordproximitynetwork}
\bibfield{author}{\bibinfo{person}{Magdalena Kaiser},
  \bibinfo{person}{Rishiraj~Saha Roy}, {and} \bibinfo{person}{Gerhard Weikum}.}
  \bibinfo{year}{2020}\natexlab{}.
\newblock \showarticletitle{Conversational Question Answering over Passages by
  Leveraging Word Proximity Networks}.
\newblock \bibinfo{journal}{\emph{arXiv preprint arXiv:2004.13117}}
  (\bibinfo{year}{2020}).
\newblock


\bibitem[\protect\citeauthoryear{Kelly, Gyllstrom, and Bailey}{Kelly
  et~al\mbox{.}}{2009}]%
        {querysuggestion3}
\bibfield{author}{\bibinfo{person}{Diane Kelly}, \bibinfo{person}{Karl
  Gyllstrom}, {and} \bibinfo{person}{Earl~W Bailey}.}
  \bibinfo{year}{2009}\natexlab{}.
\newblock \showarticletitle{A comparison of query and term suggestion features
  for interactive searching}. In \bibinfo{booktitle}{\emph{Proceedings of the
  32nd international ACM SIGIR conference on Research and development in
  information retrieval}}. \bibinfo{pages}{371--378}.
\newblock


\bibitem[\protect\citeauthoryear{Lowe, Pow, Serban, and Pineau}{Lowe
  et~al\mbox{.}}{2015}]%
        {udcdataset}
\bibfield{author}{\bibinfo{person}{Ryan Lowe}, \bibinfo{person}{Nissan Pow},
  \bibinfo{person}{Iulian Serban}, {and} \bibinfo{person}{Joelle Pineau}.}
  \bibinfo{year}{2015}\natexlab{}.
\newblock \showarticletitle{The ubuntu dialogue corpus: A large dataset for
  research in unstructured multi-turn dialogue systems}.
\newblock \bibinfo{journal}{\emph{arXiv preprint arXiv:1506.08909}}
  (\bibinfo{year}{2015}).
\newblock


\bibitem[\protect\citeauthoryear{Mazar{\'e}, Humeau, Raison, and
  Bordes}{Mazar{\'e} et~al\mbox{.}}{2018}]%
        {redditdataset}
\bibfield{author}{\bibinfo{person}{Pierre-Emmanuel Mazar{\'e}},
  \bibinfo{person}{Samuel Humeau}, \bibinfo{person}{Martin Raison}, {and}
  \bibinfo{person}{Antoine Bordes}.} \bibinfo{year}{2018}\natexlab{}.
\newblock \showarticletitle{Training millions of personalized dialogue agents}.
\newblock \bibinfo{journal}{\emph{arXiv preprint arXiv:1809.01984}}
  (\bibinfo{year}{2018}).
\newblock


\bibitem[\protect\citeauthoryear{Mei, Zhou, and Church}{Mei
  et~al\mbox{.}}{2008}]%
        {querysuggestion1}
\bibfield{author}{\bibinfo{person}{Qiaozhu Mei}, \bibinfo{person}{Dengyong
  Zhou}, {and} \bibinfo{person}{Kenneth Church}.}
  \bibinfo{year}{2008}\natexlab{}.
\newblock \showarticletitle{Query suggestion using hitting time}. In
  \bibinfo{booktitle}{\emph{Proceedings of the 17th ACM conference on
  Information and knowledge management}}. \bibinfo{pages}{469--478}.
\newblock


\bibitem[\protect\citeauthoryear{Mele, Muntean, Nardini, Perego, Tonellotto,
  and Frieder}{Mele et~al\mbox{.}}{2020}]%
        {topicpropagation}
\bibfield{author}{\bibinfo{person}{Ida Mele}, \bibinfo{person}{Cristina~Ioana
  Muntean}, \bibinfo{person}{Franco~Maria Nardini}, \bibinfo{person}{Raffaele
  Perego}, \bibinfo{person}{Nicola Tonellotto}, {and} \bibinfo{person}{Ophir
  Frieder}.} \bibinfo{year}{2020}\natexlab{}.
\newblock \showarticletitle{Topic Propagation in Conversational Search}.
\newblock \bibinfo{journal}{\emph{arXiv preprint arXiv:2004.14054}}
  (\bibinfo{year}{2020}).
\newblock


\bibitem[\protect\citeauthoryear{Moon, Shah, Kumar, and Subba}{Moon
  et~al\mbox{.}}{2019}]%
        {opendialkgdataset}
\bibfield{author}{\bibinfo{person}{Seungwhan Moon}, \bibinfo{person}{Pararth
  Shah}, \bibinfo{person}{Anuj Kumar}, {and} \bibinfo{person}{Rajen Subba}.}
  \bibinfo{year}{2019}\natexlab{}.
\newblock \showarticletitle{OpenDialKG: Explainable Conversational Reasoning
  with Attention-based Walks over Knowledge Graphs}. In
  \bibinfo{booktitle}{\emph{Proceedings of the 57th Annual Meeting of the
  Association for Computational Linguistics}}.
\newblock


\bibitem[\protect\citeauthoryear{Qu, Yang, Chen, Qiu, Croft, and Iyyer}{Qu
  et~al\mbox{.}}{2020}]%
        {qu2020open}
\bibfield{author}{\bibinfo{person}{Chen Qu}, \bibinfo{person}{Liu Yang},
  \bibinfo{person}{Cen Chen}, \bibinfo{person}{Minghui Qiu},
  \bibinfo{person}{W~Bruce Croft}, {and} \bibinfo{person}{Mohit Iyyer}.}
  \bibinfo{year}{2020}\natexlab{}.
\newblock \showarticletitle{Open-Retrieval Conversational Question Answering}.
\newblock \bibinfo{journal}{\emph{arXiv preprint arXiv:2005.11364}}
  (\bibinfo{year}{2020}).
\newblock


\bibitem[\protect\citeauthoryear{Qu, Yang, Croft, Trippas, Zhang, and Qiu}{Qu
  et~al\mbox{.}}{2018}]%
        {msdialogintent}
\bibfield{author}{\bibinfo{person}{C. Qu}, \bibinfo{person}{L. Yang},
  \bibinfo{person}{W.~B. Croft}, \bibinfo{person}{J. Trippas},
  \bibinfo{person}{Y. Zhang}, {and} \bibinfo{person}{M. Qiu}.}
  \bibinfo{year}{2018}\natexlab{}.
\newblock \showarticletitle{Analyzing and Characterizing User Intent in
  Information-seeking Conversations.}. In \bibinfo{booktitle}{\emph{SIGIR
  '18}}.
\newblock


\bibitem[\protect\citeauthoryear{Qu, Yang, Croft, Zhang, Trippas, and Qiu}{Qu
  et~al\mbox{.}}{2019}]%
        {msdialogintent2}
\bibfield{author}{\bibinfo{person}{C. Qu}, \bibinfo{person}{L. Yang},
  \bibinfo{person}{W.~B. Croft}, \bibinfo{person}{Y. Zhang},
  \bibinfo{person}{J. Trippas}, {and} \bibinfo{person}{M. Qiu}.}
  \bibinfo{year}{2019}\natexlab{}.
\newblock \showarticletitle{User Intent Prediction in Information-seeking
  Conversations}. In \bibinfo{booktitle}{\emph{CHIIR '19}}.
\newblock


\bibitem[\protect\citeauthoryear{Radlinski and Craswell}{Radlinski and
  Craswell}{2017}]%
        {framework2017}
\bibfield{author}{\bibinfo{person}{Filip Radlinski} {and} \bibinfo{person}{Nick
  Craswell}.} \bibinfo{year}{2017}\natexlab{}.
\newblock \showarticletitle{A theoretical framework for conversational search}.
  In \bibinfo{booktitle}{\emph{Proceedings of the 2017 conference on conference
  human information interaction and retrieval}}. \bibinfo{pages}{117--126}.
\newblock


\bibitem[\protect\citeauthoryear{Rao and Daum{\'e}~III}{Rao and
  Daum{\'e}~III}{2018}]%
        {raodaume2018}
\bibfield{author}{\bibinfo{person}{Sudha Rao} {and} \bibinfo{person}{Hal
  Daum{\'e}~III}.} \bibinfo{year}{2018}\natexlab{}.
\newblock \showarticletitle{Learning to ask good questions: Ranking
  clarification questions using neural expected value of perfect information}.
\newblock \bibinfo{journal}{\emph{arXiv preprint arXiv:1805.04655}}
  (\bibinfo{year}{2018}).
\newblock


\bibitem[\protect\citeauthoryear{Rao and III}{Rao and III}{2019}]%
        {raodaume2019}
\bibfield{author}{\bibinfo{person}{Sudha Rao} {and}
  \bibinfo{person}{Hal~Daum{\'{e}} III}.} \bibinfo{year}{2019}\natexlab{}.
\newblock \showarticletitle{Answer-based Adversarial Training for Generating
  Clarification Questions}.
\newblock \bibinfo{journal}{\emph{CoRR}}  \bibinfo{volume}{abs/1904.02281}
  (\bibinfo{year}{2019}).
\newblock
\showeprint[arxiv]{1904.02281}
\urldef\tempurl%
\url{http://arxiv.org/abs/1904.02281}
\showURL{%
\tempurl}


\bibitem[\protect\citeauthoryear{Rosset, Xiong, Song, Campos, Craswell, Tiwary,
  and Bennett}{Rosset et~al\mbox{.}}{2020}]%
        {rosset2020leading}
\bibfield{author}{\bibinfo{person}{Corbin Rosset}, \bibinfo{person}{Chenyan
  Xiong}, \bibinfo{person}{Xia Song}, \bibinfo{person}{Daniel Campos},
  \bibinfo{person}{Nick Craswell}, \bibinfo{person}{Saurabh Tiwary}, {and}
  \bibinfo{person}{Paul Bennett}.} \bibinfo{year}{2020}\natexlab{}.
\newblock \showarticletitle{Leading Conversational Search by Suggesting Useful
  Questions}. In \bibinfo{booktitle}{\emph{Proceedings of The Web Conference
  2020}}. \bibinfo{pages}{1160--1170}.
\newblock


\bibitem[\protect\citeauthoryear{Sordoni, Bengio, Vahabi, Lioma, Grue~Simonsen,
  and Nie}{Sordoni et~al\mbox{.}}{2015}]%
        {querysuggestion4}
\bibfield{author}{\bibinfo{person}{Alessandro Sordoni}, \bibinfo{person}{Yoshua
  Bengio}, \bibinfo{person}{Hossein Vahabi}, \bibinfo{person}{Christina Lioma},
  \bibinfo{person}{Jakob Grue~Simonsen}, {and} \bibinfo{person}{Jian-Yun Nie}.}
  \bibinfo{year}{2015}\natexlab{}.
\newblock \showarticletitle{A hierarchical recurrent encoder-decoder for
  generative context-aware query suggestion}. In
  \bibinfo{booktitle}{\emph{Proceedings of the 24th ACM International on
  Conference on Information and Knowledge Management}}.
  \bibinfo{pages}{553--562}.
\newblock


\bibitem[\protect\citeauthoryear{Su, Guo, Fan, Lan, and Cheng}{Su
  et~al\mbox{.}}{2019}]%
        {surisk}
\bibfield{author}{\bibinfo{person}{Lixin Su}, \bibinfo{person}{Jiafeng Guo},
  \bibinfo{person}{Yixing Fan}, \bibinfo{person}{Yanyan Lan}, {and}
  \bibinfo{person}{Xueqi Cheng}.} \bibinfo{year}{2019}\natexlab{}.
\newblock \showarticletitle{Controlling Risk of Web Question Answering}.
\newblock \bibinfo{journal}{\emph{CoRR}}  \bibinfo{volume}{abs/1905.10077}
  (\bibinfo{year}{2019}).
\newblock
\showeprint[arxiv]{1905.10077}
\urldef\tempurl%
\url{http://arxiv.org/abs/1905.10077}
\showURL{%
\tempurl}


\bibitem[\protect\citeauthoryear{Sun and Zhang}{Sun and Zhang}{2018}]%
        {conversationalrecommend}
\bibfield{author}{\bibinfo{person}{Yueming Sun} {and} \bibinfo{person}{Yi
  Zhang}.} \bibinfo{year}{2018}\natexlab{}.
\newblock \showarticletitle{Conversational recommender system}. In
  \bibinfo{booktitle}{\emph{The 41st International ACM SIGIR Conference on
  Research \& Development in Information Retrieval}}.
  \bibinfo{pages}{235--244}.
\newblock


\bibitem[\protect\citeauthoryear{Trienes and Balog}{Trienes and Balog}{2019}]%
        {cqidentify}
\bibfield{author}{\bibinfo{person}{Jan Trienes} {and}
  \bibinfo{person}{Krisztian Balog}.} \bibinfo{year}{2019}\natexlab{}.
\newblock \showarticletitle{Identifying unclear questions in community question
  answering websites}. In \bibinfo{booktitle}{\emph{European Conference on
  Information Retrieval}}. Springer, \bibinfo{pages}{276--289}.
\newblock


\bibitem[\protect\citeauthoryear{Voskarides, Li, Ren, Kanoulas, and
  de~Rijke}{Voskarides et~al\mbox{.}}{2020}]%
        {queryresolution}
\bibfield{author}{\bibinfo{person}{Nikos Voskarides}, \bibinfo{person}{Dan Li},
  \bibinfo{person}{Pengjie Ren}, \bibinfo{person}{Evangelos Kanoulas}, {and}
  \bibinfo{person}{Maarten de Rijke}.} \bibinfo{year}{2020}\natexlab{}.
\newblock \showarticletitle{Query Resolution for Conversational Search with
  Limited Supervision}.
\newblock \bibinfo{journal}{\emph{arXiv preprint arXiv:2005.11723}}
  (\bibinfo{year}{2020}).
\newblock


\bibitem[\protect\citeauthoryear{Wang and Ai}{Wang and Ai}{2021}]%
        {wang2021controlling}
\bibfield{author}{\bibinfo{person}{Zhenduo Wang} {and} \bibinfo{person}{Qingyao
  Ai}.} \bibinfo{year}{2021}\natexlab{}.
\newblock \showarticletitle{Controlling the Risk of Conversational Search via
  Reinforcement Learning}. In \bibinfo{booktitle}{\emph{Proceedings of the Web
  Conference 2021}}. \bibinfo{pages}{1968--1977}.
\newblock


\bibitem[\protect\citeauthoryear{Xu, Wang, Tang, Duan, Yang, Zeng, Zhou, and
  Sun}{Xu et~al\mbox{.}}{2019}]%
        {xuasking2019}
\bibfield{author}{\bibinfo{person}{Jingjing Xu}, \bibinfo{person}{Yuechen
  Wang}, \bibinfo{person}{Duyu Tang}, \bibinfo{person}{Nan Duan},
  \bibinfo{person}{Pengcheng Yang}, \bibinfo{person}{Qi Zeng},
  \bibinfo{person}{Ming Zhou}, {and} \bibinfo{person}{Xu Sun}.}
  \bibinfo{year}{2019}\natexlab{}.
\newblock \showarticletitle{Asking Clarification Questions in Knowledge-Based
  Question Answering}. In \bibinfo{booktitle}{\emph{Proceedings of the 2019
  Conference on Empirical Methods in Natural Language Processing and the 9th
  International Joint Conference on Natural Language Processing
  (EMNLP-IJCNLP)}}. \bibinfo{publisher}{Association for Computational
  Linguistics}, \bibinfo{address}{Hong Kong, China},
  \bibinfo{pages}{1618--1629}.
\newblock
\urldef\tempurl%
\url{https://doi.org/10.18653/v1/D19-1172}
\showDOI{\tempurl}


\bibitem[\protect\citeauthoryear{Yang, Qiu, Qu, Guo, Zhang, Croft, Huang, and
  Chen}{Yang et~al\mbox{.}}{2018}]%
        {msdialogrank}
\bibfield{author}{\bibinfo{person}{L. Yang}, \bibinfo{person}{M. Qiu},
  \bibinfo{person}{C. Qu}, \bibinfo{person}{J. Guo}, \bibinfo{person}{Y.
  Zhang}, \bibinfo{person}{W.~B. Croft}, \bibinfo{person}{J. Huang}, {and}
  \bibinfo{person}{H. Chen}.} \bibinfo{year}{2018}\natexlab{}.
\newblock \showarticletitle{Response Ranking with Deep Matching Networks and
  External Knowledge in Information-seeking Conversation Systems}. In
  \bibinfo{booktitle}{\emph{SIGIR '18}}.
\newblock


\bibitem[\protect\citeauthoryear{Yang, Zamani, Zhang, Guo, and Croft}{Yang
  et~al\mbox{.}}{2017}]%
        {yangnextquestion}
\bibfield{author}{\bibinfo{person}{Liu Yang}, \bibinfo{person}{Hamed Zamani},
  \bibinfo{person}{Yongfeng Zhang}, \bibinfo{person}{Jiafeng Guo}, {and}
  \bibinfo{person}{W~Bruce Croft}.} \bibinfo{year}{2017}\natexlab{}.
\newblock \showarticletitle{Neural matching models for question retrieval and
  next question prediction in conversation}.
\newblock \bibinfo{journal}{\emph{arXiv preprint arXiv:1707.05409}}
  (\bibinfo{year}{2017}).
\newblock


\bibitem[\protect\citeauthoryear{Zamani, Dumais, Craswell, Bennett, and
  Lueck}{Zamani et~al\mbox{.}}{2020a}]%
        {zamania}
\bibfield{author}{\bibinfo{person}{Hamed Zamani}, \bibinfo{person}{Susan
  Dumais}, \bibinfo{person}{Nick Craswell}, \bibinfo{person}{Paul Bennett},
  {and} \bibinfo{person}{Gord Lueck}.} \bibinfo{year}{2020}\natexlab{a}.
\newblock \showarticletitle{Generating clarifying questions for information
  retrieval}. In \bibinfo{booktitle}{\emph{Proceedings of The Web Conference
  2020}}. \bibinfo{pages}{418--428}.
\newblock


\bibitem[\protect\citeauthoryear{Zamani, Lueck, Chen, Quispe, Luu, and
  Craswell}{Zamani et~al\mbox{.}}{2020b}]%
        {zamani2020mimics}
\bibfield{author}{\bibinfo{person}{Hamed Zamani}, \bibinfo{person}{Gord Lueck},
  \bibinfo{person}{Everest Chen}, \bibinfo{person}{Rodolfo Quispe},
  \bibinfo{person}{Flint Luu}, {and} \bibinfo{person}{Nick Craswell}.}
  \bibinfo{year}{2020}\natexlab{b}.
\newblock \showarticletitle{Mimics: A large-scale data collection for search
  clarification}.
\newblock \bibinfo{journal}{\emph{arXiv preprint arXiv:2006.10174}}
  (\bibinfo{year}{2020}).
\newblock


\bibitem[\protect\citeauthoryear{Zamani, Mitra, Chen, Lueck, Diaz, Bennett,
  Craswell, and Dumais}{Zamani et~al\mbox{.}}{2020c}]%
        {zamani2020engagement}
\bibfield{author}{\bibinfo{person}{Hamed Zamani}, \bibinfo{person}{Bhaskar
  Mitra}, \bibinfo{person}{Everest Chen}, \bibinfo{person}{Gord Lueck},
  \bibinfo{person}{Fernando Diaz}, \bibinfo{person}{Paul~N Bennett},
  \bibinfo{person}{Nick Craswell}, {and} \bibinfo{person}{Susan~T Dumais}.}
  \bibinfo{year}{2020}\natexlab{c}.
\newblock \showarticletitle{Analyzing and Learning from User Interactions for
  Search Clarification}. In \bibinfo{booktitle}{\emph{Proceedings of the 43rd
  International ACM SIGIR Conference on Research and Development in Information
  Retrieval}}. \bibinfo{pages}{1181--1190}.
\newblock


\bibitem[\protect\citeauthoryear{Zhang, Chen, Ai, Yang, and Croft}{Zhang
  et~al\mbox{.}}{2018}]%
        {systemaskuserrespond}
\bibfield{author}{\bibinfo{person}{Yongfeng Zhang}, \bibinfo{person}{Xu Chen},
  \bibinfo{person}{Qingyao Ai}, \bibinfo{person}{Liu Yang}, {and}
  \bibinfo{person}{W~Bruce Croft}.} \bibinfo{year}{2018}\natexlab{}.
\newblock \showarticletitle{Towards conversational search and recommendation:
  System ask, user respond}. In \bibinfo{booktitle}{\emph{Proceedings of the
  27th ACM International Conference on Information and Knowledge Management}}.
  \bibinfo{pages}{177--186}.
\newblock


\end{thebibliography}


\end{document}